\documentclass[twoside,onecolumn]{IEEEtran}
\usepackage{amssymb}
\usepackage{graphicx}

\usepackage{amsmath}

\begin{document}

\newtheorem{corollary}{Corollary}
\newtheorem{definition}{Definition}
\newtheorem{example}{Example}
\newtheorem{lemma}{Lemma}
\newtheorem{theorem}{Theorem}
\renewcommand{\choose}[2]{{{#1}\atopwithdelims(){#2}}}

\title{An Equivalence of Entanglement-Assisted Transformation and Multiple-Copy Entanglement Transformation\thanks{This work
was partly supported by the National Foundation of Natural
Sciences of China (Grant Nos: 60321002, 60496321 and 60305005) and
the Key Grant Project of Chinese Ministry of Education (Grant No:
10403). The work of Runyao Duan was also partly supported by a PhD
Student Creative Foundation of Tsinghua University (Grant No:
05242003) }}
\author{Runyao Duan,\ \ \ Yuan Feng,\ \ \ and Mingsheng Ying\thanks{ The authors are with
the State Key Laboratory of Intelligent Technology and Systems,
Department of Computer Science and Technology, Tsinghua
University, Beijing, China 100084. E-mails:
dry02@mails.tsinghua.edu.cn (Runyao Duan), feng-y@tsinghua.edu.cn
(Yuan Feng), and yingmsh@tsinghua.edu.cn (Mingsheng Ying)}}
\date{\today}
\maketitle

\begin{abstract}
We examine the powers of entanglement-assisted transformation and
multiple-copy entanglement transformation. First, we find a
sufficient condition of  when a given catalyst is useful in
producing another specific target state. As an application of this
condition, for any non-maximally entangled bipartite pure state
and  any integer $n$ not less than $4$, we are able to explicitly
construct a set of $n\times n$ quantum states which can be
produced  by using the given state as a catalyst. Second, we prove
that for any positive integer $k$, entanglement-assisted
transformation with $k\times k$-dimensional  catalysts is useful
in producing a target state if and only if multiple-copy
entanglement transformation with $k$ copies of state is useful in
producing the same target. Moreover, a necessary and sufficient
condition for both of them is obtained in terms of the Schmidt
coefficients of the target.  This equivalence of
entanglement-assisted transformation and multiple-copy
entanglement transformation implies many interesting properties of
entanglement transformation. Furthermore, these results are
generalized to the case of probabilistic entanglement
transformations.
\smallskip\

\end{abstract}

\section{Introduction}
Quantum entanglement has been realized by the quantum information
processing community as a valuable resource, and it has been
widely used in quantum cryptography \cite{BB84}, quantum
superdense coding \cite{BS92}, and quantum teleportation
\cite{BBC+93}. A considerable amount of literature has been
devoted to the study of quantum entanglement, and many interesting
results have been reported. Nevertheless, some fundamental
problems related to quantum entanglement are still open.
Consequently, it remains the subject of interest at present after
years of investigations, see \cite{M00} for an excellent
exposition.

Since quantum entanglement often exists between different
subsystems of a composite system shared by spatially separated
parties, a natural constraint on the manipulation of entanglement
is that the separated parties are only allowed to perform quantum
operations on their own subsystems and to communicate to each
other classically. The manipulations complying with such a
constraint are called LOCC transformations. Using this restricted
set of transformations, the parties are usually required to
optimally manipulate the nonlocal resource contained in the
initial entangled state.

A central problem about quantum entanglement is thus to find the
condition of when a given entangled state can be transformed into
another one via LOCC.  Bennett and his collaborators \cite{BBPS96}
have made a significant progress in attacking this challenging
problem for the asymptotic case.  The first important step of
entanglement transformation in finite regime was made by Nielsen
in \cite{NI99}, where he presented the condition of two bipartite
entangled pure states $|\psi\rangle$ and $|\varphi\rangle$ with
the property that $|\psi\rangle$ can be locally converted into
$|\varphi\rangle$ deterministically. More precisely, let
$|\psi\rangle=\sum_{i=1}^n\sqrt{\alpha_i}|i_A\rangle|i_B\rangle$
and
$|\varphi\rangle=\sum_{i=1}^n\sqrt{\beta_i}|i_A\rangle|i_B\rangle$
be pure bipartite entangled states with ordered Schmidt
coefficient vectors $\lambda_\psi=(\alpha_1, \alpha_2, \cdots,
\alpha_n)$ and $\lambda_{\varphi}=(\beta_1,\beta_2,\cdots,
\beta_n)$, where $\alpha_1\geq \alpha_2\geq \cdots \geq
\alpha_n>0$ and $\beta_1\geq \beta_2\geq \cdots\geq \beta_n\geq
0$. Then Nielsen proved that the transformation
$|\psi\rangle\rightarrow |\varphi\rangle$ can be achieved with
certainty by LOCC if and only if $\lambda_\psi\prec
\lambda_\varphi$.  Here the symbol `$\prec $' denotes majorization
relation, and $\lambda_\psi$ is majorized by $\lambda_\varphi$ if
the following relations hold
\begin{equation}\label{nielsen}
e_l(\lambda_\psi)\leq e_l(\lambda_\varphi) {\rm \ \ \ for\ any\ \
\ }1\leq l< n,
\end{equation}
where $e_l(\lambda_\psi)=\sum_{i=1}^l\alpha_i$, i.e., the sum of
$l$ largest Schmidt coefficients of $|\psi\rangle$ (Note here by
the normalization condition, we have
$e_n(\lambda_\psi)=e_n(\lambda_\varphi)=1$). It is worth noting
that majorization relation has a natural symmetry.  Specifically,
if $E_l(\lambda_\psi)$ denotes the abbreviation of the sum of  $l$
least Schmidt coefficients of $|\psi\rangle$, then
$\lambda_{\psi}\prec \lambda_{\varphi}$ can be simply restated as
$E_l(\lambda_\psi)\geq E_l(\lambda_\varphi)$ for any $1\leq l< n$.
To see this, we only need to notice that
$e_l(\lambda_\psi)+E_{n-l}(\lambda_\varphi)=1$ holds for every
$l=1,\cdots, n-1$. We simply say that Nielsen's theorem occupies a
symmetric property.

It is well-known in linear algebra that majorization relation
$\prec$ is not a total ordering. Thus, Nielsen's theorem in fact
implies that there exist two incomparable entangled states
$|\psi\rangle$ and $|\varphi\rangle$ in the sense that neither
$|\psi\rangle \rightarrow |\varphi\rangle$ nor $|\varphi\rangle
\rightarrow |\psi\rangle$ can be realized with certainty under
LOCC.  To deal with the transformations between  incomparable
states, Vidal \cite{Vidal99} generalized Nielsen's result in a
probabilistic manner and found an explicit expression of the
maximal conversion probability for $|\psi\rangle \rightarrow
|\varphi\rangle$ under LOCC. To be more specific, let $P_{\rm
max}(|\psi\rangle \rightarrow |\varphi\rangle)$ denote the maximal
conversion probability of transforming $|\psi\rangle$ into
$|\varphi\rangle$ by LOCC. Then it was shown that
\begin{equation}\label{vidal}
P_{\rm max}(|\psi\rangle \rightarrow |\varphi\rangle)=\min_{1\leq
l\leq n} \frac{E_l(\lambda_{\psi})}{E_l(\lambda_{\varphi})}.
\end{equation}
Equivalently, we have $P_{\rm max}(|\psi\rangle\rightarrow
|\varphi\rangle)\geq p$ if and only if $E_l(\lambda_\psi)\geq
pE_l(\lambda_\varphi)$ for each $l=1,\cdots, n$. It is obvious in
this case we cannot replace $E_l(.)$ by $e_l(.)$ and then reverse
the order of inequalities. Intuitively, we may say that the
natural symmetry occupied by Nielsen's theorem is lost in the
Vidal's theorem.

Shortly after Nielsen's work, a startling phenomenon of
entanglement, namely, entanglement catalysis, or ELOCC, was
discovered by Jonathan and Plenio~\cite{JP99}. They demonstrated
by examples  that sometimes one may use an entangled state
$|\phi\rangle$, known as a catalyst, to make an impossible
transformation $|\psi\rangle \rightarrow |\varphi\rangle$
possible.  A concrete example is as follows. Take
$|\psi\rangle=\sqrt{0.4}|00\rangle+\sqrt{0.4}|11\rangle+\sqrt{0.1}|22\rangle+\sqrt{0.1}|33\rangle$
and
$|\varphi\rangle=\sqrt{0.5}|00\rangle+\sqrt{0.25}|11\rangle+\sqrt{0.25}|22\rangle$.
We know that $|\psi\rangle\nrightarrow |\varphi\rangle$ under
LOCC.  However, if another entangled state
$|\phi\rangle=\sqrt{0.6}|44\rangle+\sqrt{0.4}|55\rangle$ is
introduced, then the transformation
$|\psi\rangle\otimes|\phi\rangle \rightarrow
|\varphi\rangle\otimes|\phi\rangle$ can be realized with certainty
because $\lambda_{\psi\otimes \phi}\prec \lambda_{\varphi\otimes
\phi}$. The role of the state $|\phi\rangle$ in this
transformation is similar to a catalyst in a chemical process
since it can help entanglement transformation process without
being consumed. In the same paper, Jonathan and Plenio also showed
that the use of catalyst can improve the maximal conversion
probability when the transformation cannot be realized with
certainty even with the help of a catalyst.

Recently, Bandyopadhyay $et\ al$ \cite{SRS02} found another
interesting phenomenon:  sometimes multiple copies of source state
may be transformed into the same number of copies of target state
although the transformation cannot happen for a single copy. Such
a phenomenon is called `nonasymptotic bipartite pure-state
entanglement transformation' in \cite{SRS02}. More intuitively, we
call this phenomenon `multiple-copy entanglement transformation',
or MLOCC for short. Take the above states $|\psi\rangle$ and
$|\varphi\rangle$ as an example. It is not difficult to check that
the transformation $|\psi\rangle^{\otimes 3}\rightarrow
|\varphi\rangle^{\otimes 3}$ occurs with certainty by Nielsen's
theorem. That is, when Alice and Bob prepare three copies of
$|\psi\rangle$ instead of just a single one, they can transform
these three copies all together into three copies of
$|\varphi\rangle$ by LOCC. This simple example means that the
effect of catalyst can, at least in the above situation, be
implemented by preparing more copies of the original state and
transforming these copies together. Besides some concrete examples
of MLOCC, various theoretical properties of MLOCC were also
investigated in \cite{SRS02}.

After \cite{JP99, SRS02}, due to the great importance of
entanglement transformation in quantum information processing, a
considerable number of researches were done to investigate the
mechanism beyond entanglement catalysis and multiple-copy
entanglement transforamtion. For example, in \cite{DK01}, Daftuar
and Klimesh carefully examined the mathematical structure of
entanglement catalysis. They showed that any non-maximally
bipartite entangled   pure state can serve as quantum catalyst for
some entanglement transformation. Especially, the relationship
between entanglement catalysis and multiple-copy entanglement
transformation has been thoroughly studied by the authors in
\cite{DFLY04}. It was proved that any multiple-copy entanglement
transformation can be implemented by a suitable
entanglement-assisted transformation.  Another essential
connection between entanglement-assisted transformation  and
multiple-copy entanglement transformation was also presented in
\cite{DFLY04}. Indeed, the equivalence between the possibility of
implementing an entanglement transformation in producing a  given
target by ELOCC and the one by MLOCC was observed.

In this paper we examine the powers of entanglement-assisted
transformation and multiple-copy entanglement transformation from
some new angles. The first problem we consider here is the
usefulness of a given catalyst in producing a target state. To be
concise, we say that a catalyst $|\phi\rangle$ is useful in
producing a target \ $|\varphi\rangle$ if there exists some pure
state $|\psi\rangle$ with the same dimension as $|\varphi\rangle$
such that $|\psi\rangle\otimes |\phi\rangle\rightarrow
|\varphi\rangle\otimes |\phi\rangle$ can be achieved with
certainty by LOCC while $|\psi\rangle$ cannot be transformed to
$|\varphi\rangle$ directly. To solve the problem of usefulness of
catalyst, two simple but useful mathematical apparatuses are
introduced, namely, local uniformity and global uniformity. They
enable us to give a sufficient condition of whether a catalyst
$|\phi\rangle$ is useful in producing $|\varphi\rangle$(Theorem
\ref{catalystuseful}). More importantly, this condition is
operational and it determines all catalyst states with the minimal
dimension. Thus, it is very useful in practice.

The second problem that we consider in the present paper is  to
determine whether there exists some $k\times k$ catalyst
$|\phi\rangle$ which is useful in producing  $|\varphi\rangle$,
where $k\geq 2$ is a given dimension. This problem is slightly
different from the previous one. The major difference is  the
catalyst state in the first problem is specified while in the
current problem only the dimension of the catalyst state is fixed.
If such a $k\times k$ catalyst state does exist, we simply say
that $k$-ELOCC is useful in producing $|\varphi\rangle$.

A corresponding problem occurs when we consider multiple-copy
entanglement transformation. If there exists some $n\times n$
state $|\psi\rangle$ such that $|\psi\rangle^{\otimes
k}\rightarrow |\varphi\rangle^{\otimes k}$ can be achieved with
certainty while $|\psi\rangle\nrightarrow |\varphi\rangle$ under
LOCC, then we say that $k$-MLOCC is useful in producing
$|\varphi\rangle$. Thus the third  problem may be more precisely
stated as follows: for a given state $|\varphi\rangle$ and a
positive integer $k>1$, decide  whether $k$-MLOCC is  useful in
producing $|\varphi\rangle$.

The above two problems are concerned respectively with $k$-ELOCC
and $k$-MLOCC, and it seems that they are irrelevant. To our
surprise, we find that these two problems are equivalent. Indeed,
we show that for any bipartite entangled state $|\varphi\rangle$
and positive integer $k$, $k$-ELOCC is useful in producing
$|\varphi\rangle$ if and only if $k$-MLOCC is useful in producing
$|\varphi\rangle$. Furthermore, a necessary and sufficient
condition for both of them is obtained in terms of the Schmidt
coefficients of $|\varphi\rangle$ (Theorem \ref{kmelocc}). As a
simple corollary, we are also able to prove a similar equivalence
between ELOCC and MLOCC for the case in which the dimension of
catalysts (or the number of copies) is not fixed. This complements
further the results of \cite{DK01} and \cite{DFLY04} mentioned
above.

The previous results are obtained in the deterministic case. We
are able to solve the above problems for the case of probabilistic
transformations too. However, our results show that some
properties of probabilistic ELOCC and MLOCC transformations are
quite different from those of their deterministic counterparts
(Theorems \ref{characterizationforcatalystusefulprob} and
\ref{kemloccprob}). This is somewhat surprising. We argue that
this phenomenon is deeply related to the difference between the
mathematical structures of deterministic entanglement
transformations and probabilistic entanglement transformations,
which are characterized by Nielsen's theorem and Vidal's theorem,
respectively. As pointed out above, Nielsen's Theorem enjoys a
natural symmetry, but this symmetry is lost in Vidal's theorem.

We organize the rest of this paper as follows. We state our main
results in Section II. Some direct implications are also pointed
out there. In Section III, we give some interesting applications
of the main results. In particular, two conjectures of Nielsen
about entanglement catalysis are addressed in detail. From Section
IV on, we present the proofs of the main results. In Section IV,
we give some lemmas which are needed in these proofs of the main
results. The rest several sections completes the proofs. Theorem
\ref{catalystuseful} will be proved  in Section V.  In Section VI,
we present the proof of Theorem \ref{kmelocc}. Theorems
\ref{characterizationforcatalystusefulprob} and \ref{kemloccprob}
will be proved in Section VII. To keep the paper more readable, we
put the complicated proofs of some technical lemmas in the
appendices. Along with the proofs, many properties about the
mathematical structure of ELOCC and MLOCC which are independently
of interest are also presented. A brief conclusion is drawn in
Section VIII.

\section{Main results}

The purpose of this section is to state the main results. The
proofs of them are postponed to Sections V - VII. Since the
fundamental properties of a bipartite pure state under LOCC are
completely determined by its Schmidt coefficients, which can be
treated as a probability vector, we consider only probability
vectors instead of quantum states from now on. We always identify
a probability vector with the quantum state represented by it.

\subsection{Notations and Definitions}

To present the main results, we need some auxiliary notations. Let
$V^n$ denote the set of all $n$-dimensional probability vectors.
For any $x\in V^n$, the dimensionality of $x$ is often denoted by
${\rm dim}(x)$, i.e., ${\rm dim}(x)=n$.  The notation
$x^{\downarrow}$ will be used to stand for the vector which is
obtained by rearranging the components of $x$ into non-increasing
order. We use $e_l(x)$ to denote the sum of  $l$ largest
components of $x$, i.e., $e_l(x)=\sum_{i=1}^{l}x^{\downarrow}_i$.
It is obvious  that $e_l(x)$ is a  continuous function of $x$ for
each $l=1,2,\cdots, n$.

We say that $x$ is majorized by $y$, denoted by $x\prec y$, if
\begin{equation}\label{nielsentheorem} e_m(x)\leq e_m(y)
{\rm \ for\ every\ }m=1,\cdots, n-1,
\end{equation}
with equality if $m=n$. If all inequalities in
Eq.(\ref{nielsentheorem}) are strict, we say that $x$ is  strictly
majorized by $y$. The relation of strict majorization is
represented by $x\lhd y$.

Using the above notations,  Nielsen's theorem can be stated as:
$x\rightarrow y$ under LOCC if and only if  $x\prec y$.

Although we consider probability vectors only, we often omit the
normalization step for simplicity. This has no influence on the
validity of our results. We can assume that all catalyst
probability vectors have positive components because states $c$
and $c\oplus 0$ are equivalent when they are treated as catalysts.

We also assume that the components of probability vector
$x=(x_1,\cdots, x_n)$ are always  in non-increasing order, except
where otherwise stated.  We say that $x$ is a segment of another
vector $y$ if there exist $i\geq 1$ and $k\geq 0$ such that
$x=(y_i, y_{i+1},\cdots, y_{i+k})$.

Two useful quantities  named local uniformity and global
uniformity are key mathematical tools in the present paper.
Formally, we have the following definitions:

\begin{definition}\label{uniformity}\upshape
Let $x$ be an $n$-dimensional  probability vector.

1) The local uniformity of $x$ is defined by
\begin{equation}\label{localuniformity}
 l_u(x)={\rm min}\{\frac{x_{i+1}}{x_i}: 1\leq
 i<n-1\}.
\end{equation}

2) The global uniformity of $x$ is defined by
\begin{equation}\label{globaluniformity}
g_u(x)=\frac{x_n}{x_1}.
\end{equation}
\end{definition}

By the above definition, we have that both $l_u(x)$ and $g_u(x)$
are between $0$ and $1$. The above definition can be extended to
any positive vector which is not necessarily normalized.

A simple but useful relation between $l_u(x)$ and $g_u(x)$ is the
following:
\begin{equation}\label{lugu}
l_u^{n-1}(x)\leq g_u(x)\leq l_u(x),
\end{equation}
which will be used again and again.

\subsection{When is a catalyst $c$ useful in producing a target state $y$?}
For any $y\in V^n$, we write $S(y)=\{x\in V^n: x \prec y\}$.
Intuitively, $S(y)$ denotes all the probability vectors which can
be transformed into $y$ by LOCC. We also define $T(y,c)$ to be the
set of all  probability vectors that can be transformed into $y$
with $c$ as a catalyst, i.e., $T(y,c)=\{x\in V^n: x\otimes c\prec
y\otimes c\}$. In practice, an important problem is to find
catalyst states which are useful in producing  a given target
state. This is exactly the first problem that we promised to
attack in the introduction. With the notations introduced above it
can be briefly reformulated: whether $S(y)\subsetneqq T(y,c)$
holds. The following theorem gives a partial answer to this
problem. More exactly, it presents a sufficient condition under
which a given target entangled state can be  implemented by using
a given catalyst. It is worth mentioning that this condition is
operational and it is also almost necessary.

\begin{theorem}\label{catalystuseful}\upshape
Let $y\in V^n$. If a catalyst $c$ satisfies
\begin{equation}\label{usecata}
l_u(c)>{\rm max}\{\frac{y_d}{y_1},\frac{y_n}{y_{d+1}}\} {\rm \
and\ } g_u(c)<\frac{y_{d+1}}{y_d}
\end{equation}
for some $1<d<n-1$, then  $S(y)\subsetneqq T(y,c).$ Conversely, if
$S(y)\subsetneqq T(y,c)$, then there is a segment of $c$
satisfying Eq.(\ref{usecata}).
\end{theorem}

Some remarks come as follows:

\begin{enumerate}

\item If $c$ is a vector with the minimal dimension such that
$S(y)\subsetneqq T(y,c)$, then by the above theorem, $c$ should
satisfy Eq.(\ref{usecata}). In the view of this, Theorem
\ref{catalystuseful} determines all catalyst states with the
minimal dimension which are useful in producing  $y$.  Especially,
if $c$ has only two distinct non-zero components, then
Eq.(\ref{usecata}) is also necessary for $S(y)\subsetneqq T(y,c)$.

\item A direct consequence of the above theorem is that any
uniform vector  $c$ cannot serve as a catalyst for any vector $y$
since Eq.(\ref{usecata}) cannot be satisfied. Furthermore, we can
show that  any nonuniform probability vector can serve as a
quantum catalyst for uncountably many probability vectors, which
is a considerable improvement of the result proved by Daftuar and
Klimesh in \cite{DK01}. And this gives a stronger answer of
Nielsen's conjecture \cite{MAJ}, which states that any nonuniform
probability vector can potentially serve as a catalyst for some
transformation.
\end{enumerate}

The applications mentioned above will be discussed in much more
details in Section III. The proof of Theorem \ref{catalystuseful}
will be presented in Section V.

\subsection{Equivalence of ELOCC and MLOCC}

Now we  further review some  elements of entanglement catalysis
and multiple-copy entanglement transformation. For any $y\in V^n$,
let $T(y)=\{x\in V^n:x\otimes c\prec y\otimes c\ {\rm for \ some\
vector\ } c\}.$ Intuitively, $T(y)$ denotes the probability
vectors which can be transformed into $y$ by LOCC with the help of
some catalyst. We also define $M(y)$ to be the set of probability
vectors which, when provided with a finite (but large enough)
number of copies, can be transformed into the same number of $y$
under LOCC, that is, $M(y)=\{x\in V^n: x^{\otimes{k}}\prec
y^{{\otimes{k}}} \ {\rm for\ some}\ k\ \geq 1\}$. If we restrict
the number of copies used in $M(y)$ to be $k$ and the vector $c$
used as catalyst in $T(y)$ to be $k$-dimensional, then we can
define $M_k(y)$ and $T_k(y)$ similarly; namely, $M_k(y)=\{x\in
V^{n}:x^{\otimes{k}}\prec y^{\otimes{k}}\}$ and $T_k(y)=\{x\in
V^{n}:x\otimes c\prec y\otimes c\ {\rm for \ some}\ c\in V^{k}\}$.

In \cite{DFLY04}, it was shown that $T(y)=S(y)$ if and only if
$M(y)=S(y)$. This interesting result has an intuitive physical
meaning: for any quantum state $y$, if ELOCC is useless  in
producing $y$, nor has MLOCC, and vice versa. So we get an
equivalence of ELOCC and MLOCC in the sense that they are both
useful in producing the same target or both not. In the present
paper, this result will be considerably refined. More precisely,
we prove that for a specific class of entangled states, enhancing
the number of copies but not exceeding a threshold will be
useless. Furthermore, for any positive integer $k\geq 2$, we give
a complete characterization of $T_k(y)=S(y)$ in terms of
components of $y$. A similar result for the equality $M_k(y)=S(y)$
is also proved. To one's surprise, these two conditions are in
fact the same. So we find a relation that $T_k(y)=S(y)$ if and
only if $M_k(y)=S(y)$, which is much more elaborated than that
$T(y)=S(y)$ if and only if $M(y)=S(y)$, previously established in
\cite{DFLY04}. We state this main result as the following:

\begin{theorem}\label{kmelocc}\upshape
 For any $y\in V^n$, the following  are equivalent:

1. $T_k(y)=S(y)$.

2. $M_k(y)=S(y)$.

3. $y_d^k\geq y_1^{k-1}y_{d+1} {\rm \ or\ }y_{d+1}^k\leq
y_n^{k-1}y_d$ for any $1<d<n-1$.

\end{theorem}

Let us list some implications of the above theorem as follows:
\begin{enumerate}

\item In the case that $k$ tends to infinity, items 1 and 2 in the
above theorem reduce to $T(y)=S(y)$ and $M(y)=S(y)$, respectively,
and item 3 reduces to $y_d=y_1$ or $y_{d+1}=y_n$ for any
$1<d<n-1$. Hence we recover the main results in \cite{DFLY04}.

\item A careful observation carries out that in the case that
$T(y)\neq S(y)$, there can still exist some $k>1$ such that
$T_k(y)=S(y)$. That is, although catalysis is useful in producing
$y$, any quantum states  with dimension less than $k$ cannot serve
as catalysts. This also gives a solution to an open problem
addressed by Jonathan and Plenio  in \cite{JP99}, where they asked
that whether catalyst states are always more efficient as their
dimension increases.  We also note that a similar question has
also been addressed by Daftuar and Klimesh in \cite{DK01}, where
they asked whether there are some $y\in V^n$ and $k\geq1$ such
that $T_k(y)=T_{k+1}(y)$.  Theorem \ref{kmelocc} shows that  to
make some $k$-dimensional state serve as a catalyst in producing
$y$, the components of $y$ should satisfy some conditions, which
cannot always be fulfilled by any probability vectors $y$.

\item Theorem \ref{kmelocc} can certainly help us not only to
understand the limitation of entanglement catalysis and
multiple-copy entanglement transformation, but also to choose
suitable entangled states with good properties under ELOCC and
MLOCC in practical quantum information processing.

\item Theorem \ref{kmelocc} also discovers a very surprising
connection between $k$-ELOCC and $k$-MLOCC. In \cite{DFLY04}, it
was demonstrated that $M_k(y)\subseteq T_{kn^{k-1}}(y)$, but we
still do not know whether the bound $kn^{k-1}$ is tight or not. It
seems that $T_k(y)$ and $M_k(y)$ have no any connection. To check
whether $x\in T_k(y)$,  we need to consider all the
$k$-dimensional probability vectors as possible catalysts, which
form a set of the size of continuum. But to check whether $x$ is
in $M_k(y)$, only a simple calculation whether $x^{\otimes k}\prec
y^{\otimes k}$ is needed. However, Theorem \ref{kmelocc} enables
us to build up a `weak' equivalence between these two sets:
$k$-ELOCC is useful in producing $y$ if and only if $k$-MLOCC is
useful in producing $y$.
\end{enumerate}

We will discuss the  applications of Theorem \ref{kmelocc} in more
details in Section III.  The proof of Theorem \ref{kmelocc} will
be presented in Section VI.

\subsection{Probabilistic entanglement transformations}

In the previous two subsections, we are concerned with
deterministic entanglement transformations. In this subsection, we
try to solve the same problems for probabilistic entanglement
transformations. Our main results are Theorems
\ref{characterizationforcatalystusefulprob} and \ref{kemloccprob},
and they are counterparts of Theorems \ref{catalystuseful} and
\ref{kmelocc}, respectively. It is interesting to note that the
appearance of the results in this subsection are quite different
from the corresponding ones for deterministic transformations.
Indeed, it seems impossible to unify the deterministic case and
the probabilistic case in a simple way. Even  more strange, the
probabilistic case is much simpler than the deterministic case.

In \cite{FDY04}, the mathematical structure of
entanglement-assisted probabilistic transformations was thoroughly
studied. The results presented below complement well the ones
obtained in \cite{FDY04}.

To study probabilistic entanglement transformations, we need the
notion of super majorization. Let $x$ and $y$ be two
$n$-dimensional vectors. We say $x$ is super-majorized by $y$,
denoted by $x\prec^w y$, if $E_l(x)\geq E_l(y)$ for any $1\leq
l\leq n$. In the case that the sum of $x$ and $y$ are equal, i.e.,
$E_n(x)=E_n(y)$, $x\prec^w y$ reduces to $x\prec y$. We write
$x\lhd^w y$ if and only if $E_l(x)> E_l(y)$ for any $1\leq l\leq
n$.

By means of super majorization,  Vidal's theorem  can be restated
as: for any $\lambda\in (0,1)$, $P_{\rm max}(x\rightarrow y)\geq
\lambda$ if and only if $x\prec^{w} \lambda y$, where $\lambda$ is
understood as a probability threshold.

As a natural generalization of $S(y)$, we define
$S^{\lambda}(y)=\{x\in V^n: x\prec^w \lambda y\}$. Intuitively,
$S^\lambda(y)$ denotes the set of all probability vectors which
can be transformed into $y$ with a probability at least $\lambda$.
Similarly, let $T^{\lambda}(y,c)=\{x\in V^n: x\otimes
c\prec^{w}\lambda y\otimes c\}$.

The following theorem is a probabilistic counterpart of Theorem
\ref{catalystuseful}. It in fact provides a simple analytical
characterization of the catalyst states $c$ which are useful in
producing $y$ in a probabilistic manner:
\begin{theorem}\label{characterizationforcatalystusefulprob}\upshape
For any $y\in V^n$,  $S^{\lambda}(y)\subsetneqq T^{\lambda}(y,c)$
if and only if there exist $0<d<n-1$ and $1\leq i<k$ such that
\begin{equation}\label{probusecata}
l_u(c')>\frac{y_n}{y_{d+1}} {\rm\ and\
}g_u(c')<\frac{y_{d+1}}{y_d},
\end{equation}
where  $c'=(c_i,c_{i+1},\cdots, c_k)$.
\end{theorem}

To present a corresponding result with Theorem \ref{kmelocc}, we
need to generalize $T_k(y)$ and $M_k(y)$ to probabilistic
versions. Specifically,  $T^{\lambda}_k(y)=\{x\in V^n:x\otimes
c\prec^{w} \lambda y\otimes c {\rm\ for\ some\ } c\in V^k \}$
denotes the set of all quantum states that can be transformed into
$y$ with  a probability not less than $\lambda$ with the help of a
$k$-dimensional catalyst state. Let $T^{\lambda}(y)=\{x\in V^n:
x\otimes c\prec^{w} \lambda y\otimes c {\rm\  for\ some\ }c\}$. It
has a similar meaning but the dimension of catalyst state is not
fixed.  $M_k(y)$ and $M(y)$ can also be generalized to
probabilistic case as follows: $M^{\lambda}_k(y)=\{x\in V^n:
x^{\otimes k}\prec^{w} \lambda^{k} y^{\otimes k}\}$ and
$M^{\lambda}(y)=\{x\in V^n: x^{\otimes
k}\prec^{w}\lambda^{k}y^{\otimes k} {\rm\ for\ some\ } k\}$. The
physical meanings of $T_k^\lambda(y)$ and $T^\lambda(y)$ are very
clear while the definitions of $M_k^\lambda(y)$ and $M^\lambda(y)$
seem to be artificial and deserve a careful explanation. We give
an intuitive interpretation of $M_k^\lambda(y)$ and $M^\lambda(y)$
here. Noticing that for any $x\in M_k^\lambda(y)$, we have
$x^{\otimes k}\prec^w \lambda^k y^{\otimes k}$, or more
explicitly,
$$P_{\rm max}(x^{\otimes k}\rightarrow y^{\otimes k})\geq \lambda^k.$$  If the maximal conversion
probability from $x$ to $y$ by LOCC is $\lambda$, then the
right-hand side of the above inequality is just the maximal
conversion probability of transforming $x^{\otimes k}$ into
$y^{\otimes k}$ separately, that is, in a way where no collective
operations on the $k$ copies are performed. Thus the intuition
behind the above definition is that with the help of $k$-MLOCC,
the geometric average value of the probability of a single-copy
transformation is not less than $\lambda$. Similarly, $x\in
M^\lambda(y)$ means that with the help of MLOCC, the average
probability of a single-copy transformation is not less than
$\lambda$.

With these preliminaries,  we present a probabilistic counterpart
of Theorem \ref{kmelocc} in the following:
\begin{theorem}\label{kemloccprob}\upshape
For any $y\in V^n$, the following are equivalent:

1. $T^{\lambda}_k(y)=S^{\lambda}(y)$.

2. $M^{\lambda}_k(y)=S^{\lambda}(y)$.

3. $y_{d+1}^k\leq y_n^{k-1}y_d$ for any $0< d<n-1$.
\end{theorem}

We give some remarks about the above two theorems:
\begin{enumerate}

\item  It is very interesting that  the probabilistic threshold
$\lambda\in (0,1)$ is irrelevant in Theorem
\ref{characterizationforcatalystusefulprob}. In other words, for
any $c$ and $y$, whether $S^\lambda(y)\subsetneqq T^\lambda(y,c)$
does not depend on $\lambda$. Roughly speaking, this reveals a
uniformity property of entanglement-assisted probabilistic
transformations. A similar phenomenon  occurs in Theorem
\ref{kemloccprob}.

\item We may naturally expect that the deterministic case of
$\lambda=1$ can be included in Theorems
\ref{characterizationforcatalystusefulprob} and \ref{kemloccprob},
and the deterministic case and the probabilistic case can be
unified. Unfortunately, it is not the case, and Theorems
\ref{characterizationforcatalystusefulprob} and Theorem
\ref{kemloccprob} are valid only when $\lambda<1$. This fact is
deeply rooted in the symmetry of Nielsen's Theorem and the
asymmetry of Vidal's Theorem, which describe the conditions of
deterministic transformations and probabilistic transformations,
respectively.

\end{enumerate}

Some applications of these two theorems will also be presented in
the next section, and their proofs are put in Section VII.

\section{Some Applications}

\subsection{What states can be used as catalysts?}

In his lecture notes \cite{MAJ}, Nielsen conjectured that any
nonuniform probability vector can potentially serve as  catalyst
for some transformation. This conjecture was proved to be true by
Daftuar and Klimesh \cite{DK01}. In fact, they  proved that for
any nonuniform $z\in V^k$, there exist $x,y\in V^4$ such that
$x\nprec y$ but $x\otimes z\prec y\otimes z$. As an interesting
application of Theorem \ref{catalystuseful}, we further show that
any nonuniform probability vector can serve as quantum catalyst
for uncountably many probability vectors.
\begin{theorem}\label{nonuniform}\upshape
Suppose $z\in V^k$ and $z_1>z_k>0$, $n\geq 4$. There exists a
subset $A(z)$ of $V^n$  with non-zero measure relative to $V^n$,
such that for any $y\in A(z)$, $S(y)\subsetneqq T(y,z)$.
\end{theorem}

\textit{Proof.} We will explicitly construct $A(z)\subseteq V^n$
such that for any $y\in A(z)$, $S(y)\subsetneqq T(y,z)$. For a
specific $1<d<n-1$, we define $A_d(z)$ to be the set of all
probability vectors $y\in V^n$ such that
\begin{equation}\label{lvz}
l_u(z)>{\rm max}\{\frac{y_d}{y_1}, \frac{y_n}{y_{d+1}}\}{\rm\ and\
} g_u(z)<\frac{y_{d+1}}{y_d}.
\end{equation}
By Theorem \ref{catalystuseful}, it follows that $S(y)\subsetneqq
T(y,z)$. Then $A(z)$ can be defined as the union of $A_d(z)$ for
all $1<d<n-1$. It is clear that $A(z)$ has a non-zero measure
relative to $V^n$. In the case that $l_u(z)=g_u(z)$, i.e., $z$ has
only two non-zero distinct components, $A(z)$ is the set of all
probability vectors $y\in V^n$ such that $S(y)\subsetneqq T(y,z)$,
and the conclusion also follows from Theorem \ref{catalystuseful}.
This completes the proof of
Theorem \ref{nonuniform}.\hfill $\square$\\

Note that in \cite{DK01}, Daftuar{\it \ et\ al} constructed two
probability vectors
$x=(\alpha/2+\beta/4,\alpha/2+\beta/4,\beta/4,\beta/4)$ and
$y=(\alpha,\beta/2,\beta/2,0),$ where $z_1/z_k=\alpha/\beta$,
$\alpha+\beta=1$. They proved that $e_l(x\otimes z)<e_l(y\otimes
z)$ holds for each $1\leq l<4k$. So $x\otimes z\lhd y\otimes z$.
Hence they asserted that a small enough perturbation on $x$
generates the desired probability vector
$x(\epsilon)=(x_1+\epsilon, x_2,x_3,x_4-\epsilon)$ such that
$x(\epsilon)\otimes z\prec y\otimes z$ but $x(\epsilon)\nprec y$.
A trick lies in showing that $e_l(x\otimes z)<e_l(y\otimes z)$ for
any $1\leq l<4k$. To achieve this goal, they first proved that
when $l$ is even the inequality holds by considering five possible
cases according to the relationship between $l$ and $k$, and then
with a small modification they proved that when $l$ is odd the
relation $e_l(x\otimes z)<e_l(y\otimes z)$ also holds. However,
the construction of $x$ and $y$ is very artificial and the proof
is a highly skilled one. Their proof heavily depends on the
concrete instances $x$ and $y$ and cannot be generalized easily.
On the other hand, the proof presented above is a coherent one and
Theorem \ref{nonuniform} has considerably generalized the result
obtained by Daftuar{\it \ et\ al} \cite{DK01}.

To illustrate the application of Theorem \ref{nonuniform}, let us
reexamine the above example obtained by  Daftuar $et\ al$
\cite{DK01}.  We only need to show that $y\in A(z).$  Because $z$
is a nonuniform probability vector, we have $0<g_u(z)\leq
l_u(z)<1$. A routine calculation carries out that
\begin{equation}\label{dk1}
l_u(z)>{\rm max}\{\frac{y_d}{y_1}, \frac{y_n}{y_{d+1}}\} {\rm\
and\ } g_u(z)<\frac{y_{d+1}}{y_d},
\end{equation}
where $d=2$ and $n=4$. So $S(y)\subsetneqq T(y,z)$ by Theorem
\ref{catalystuseful}.  Moreover, noticing that $x_1+x_2=y_1+y_2$,
we have $x\otimes z\lhd y\otimes z$ by the proof of Theorem
\ref{catalystuseful}.

Furthermore, any $y\in V^4$ satisfying Eq.(\ref{dk1}) has  the
property such that $S(y)\subsetneqq T(y,z)$, so the example given
by Daftuar{\it \ et\ al} \cite{DK01} is only a special case.

\subsection{When  are ELOCC and MLOCC  useful?}

We turn now to give some applications of Theorem \ref{kmelocc}.

As mentioned above, we are able to recover one of the main results
in \cite{DK01} and \cite{DFLY04}. That is, for any $y\in V^n$,
$T(y)=S(y)$ if and only $M(y)=S(y)$. Moreover, an explicit
necessary and sufficient condition for the equality $T(y)=S(y)$
(and equivalently $M(y)=S(y)$) is also obtained in terms of the
components of $y$, as the following theorem states:
\begin{theorem}\label{infmelocc}\upshape
For any $y\in V^n$, the following are equivalent:

1) $T(y)=S(y)$.

2) $M(y)=S(y)$.

3) $y_d=y_1$ or $y_{d+1}=y_n$ for any $1<d<n-1$.

\end{theorem}

Although this result has been proved in \cite{DK01} and
\cite{DFLY04}, we prefer to give a completely different but much
simpler proof based on Theorem \ref{kmelocc}.

\textit{Proof.} The case of $n\leq 3$ is trivial. We assume $n\geq
4$. The equivalence of  1) and 2) is a direct consequence of
Theorem \ref{kmelocc}. We only need to show the equivalence of 1)
and 3).  Suppose that 3) does not hold, i.e., there exists
$1<d<n-1$ such that $y_d<y_1$ and $y_{d+1}>y_n$.  Then we can find
a sufficiently large $k$ such that
\begin{equation}\label{powerpower}
y_d^k<y_1^{k-1}y_{d+1} {\rm \ and\ }y_{d+1}^k> y_dy_n^{k-1},
\end{equation}
which further leads to  $S(y)\subsetneqq T_k(y)\subseteq T(y)$ by
Theorem \ref{kmelocc}. Hence  1) cannot hold.

Conversely. Suppose  $S(y)\subsetneqq T(y)$. Then  we will find
integer $k$ with $S(y)\subsetneqq T_k(y)$, which implies the
existence of $1<d<n-1$ satisfying condition
(\ref{powerpower}). Therefore 3) cannot hold. \hfill $\square$\\

Interestingly, a probabilistic  version of the above theorem is
the following:
\begin{theorem}\label{infmeloccprob}\upshape
Let $y\in V^n$ and $\lambda\in (0,1)$. Then the following are
equivalent:

1)\ $T^\lambda(y)=S^\lambda(y)$.

2)\ $M^\lambda(y)=S^\lambda(y)$.

3)\ $y_2=y_n$.
\end{theorem}

We should note that the equivalence of 1) and 3) has been proved
in \cite{FDY04}. Again, we can see that the probabilistic
threshold $\lambda$ is not involved.

We present an interesting example to illustrate the difference
between probabilistic transformations and deterministic
transformations.

\begin{example}\label{example2}\upshape
Let $y=(y_1,y_2,y_3)$ be a 3-dimensional probability vector. Then
by Theorem \ref{infmelocc}, we have that $T(y)=M(y)=S(y)$. That
is, ELOCC and MLOCC are useless for any deterministic
transformations to $y$.

On the other hand, if we take $y$ satisfying $y_2>y_3$, then by
Theorem \ref{infmeloccprob}, we have $S^\lambda (y)\subsetneqq
T^\lambda(y)$ and $S^\lambda(y)\subsetneqq M^\lambda(y)$ for any
$\lambda\in (0,1)$. That is, for any such state $y$ and
$\lambda\in (0,1)$, we can always find another state $x\in V^3$
and a catalyst $c$ such that $P_{\rm max}(x\rightarrow y)<\lambda$
but $P_{\rm max}(x\otimes c\rightarrow y\otimes c )\geq \lambda$.
Equivalently, we can find an integer $k>1$ such that $P_{\rm
max}(x^{\otimes k}\rightarrow y^{\otimes k})\geq \lambda^k$. Hence
ELOCC and MLOCC are useful for some probabilistic transformations
in producing  $y$.
\end{example}

\subsection{More applications}

In \cite{LS01}, Leung and Smolin demonstrated that $x^{\otimes
k}\prec y^{\otimes k}$ does not necessarily imply $x^{\otimes
k+1}\prec y^{\otimes k+1}$ by giving explicit instances of $x$ and
$y$, where $k$ is a positive integer not less than $2$. In other
words, for some $y\in V^n$ and $k>1$, we have $M_k(y)\nsubseteq
M_{k+1}(y)$. That is, increasing number of copies cannot always
help entanglement transformation. However, with the aid of Theorem
\ref{kmelocc}, we can prove that if $k+1$-copies of transformation
are not useful in producing the same target state, then $k$-copies
of transformation are also not useful in producing the same
target.

\begin{theorem}\label{kuseless}\upshape
For any $y\in V^n$ and $k> 1$, $M_{k+1}(y)=S(y)$ implies
$M_k(y)=S(y).$
\end{theorem}

\textit{Proof.} In fact, by Theorem \ref{kmelocc}, the condition
for $M_{k+1}(y)= S(y)$ can be rewritten as
$${\rm max\ }\{(\frac{y_d}{y_1})^{k-1},
(\frac{y_n}{y_{d+1}})^{k-1}\}\geq \frac{y_{d+1}}{y_d}.$$ Notice
that the left-hand side of the above inequality is a decreasing
function of $k$, and these inequalities still hold if we replace
$k$ with $k-1$. $M_k(y)= S(y)$ follows  immediately by using
Theorem
\ref{kmelocc} once more.\hfill $\square$\\

A direct consequence of Theorem \ref{kuseless} is the following:
\begin{corollary}\label{kless}
For any $y\in V^n$ and $k\geq 1$, $M_k(y)=S(y)$ implies
$M_l(y)=S(y){\rm \ for\ any\ } l\leq k.$
\end{corollary}

In the case of ELOCC,  if we  restrict the components of catalyst
to be positive, then whether the inclusion relation
$T_k(y)\subseteq T_{k+1}(y)$ always holds is unknown.
Nevertheless,  we can build up a corresponding result with Theorem
\ref{kuseless} as the following:
\begin{theorem}\label{ckuseless}\upshape
For any $y\in V^n$, $T_{k+1}(y)=S(y)$ implies $T_k(y)=S(y)$.
\end{theorem}

Intuitively, if $k+1$-dimensional entanglement-assisted
transformation is not usful in producing a target state, then the
$k$-dimensional entanglement transformation is also not  useful in
producing the same target.

\textit{Proof.} Similar to Theorem \ref{kuseless}, the key step
is to apply Theorem \ref{kmelocc}. We omit the details here. \hfill $\square$\\
A corresponding corollary of Theorem \ref{ckuseless} is stated as
follows:
\begin{corollary}\label{ckless}\upshape
For any $y\in V^n$ and $k\geq 1$, $T_k(y)=S(y)$ implies
$T_l(y)=S(y){\rm \ for\ any\ } l\leq k.$
\end{corollary}

A practical  application of Theorem \ref{kmelocc} is to help
finding a suitable catalyst for a given transformation
$x\rightarrow y$. In \cite{SD03}, Sun and some of us proposed a
polynomial (of $n$) time algorithm to decide whether there is some
catalyst $c\in V^k$  for the transformation from $x$ to $y$, where
$k$ is treated as a  fixed positive integer. Combining Theorem
\ref{kmelocc} with this algorithm, we can first find the minimal
$k$ such that $S(y)\subsetneqq T_k(y)$ and then use the algorithm
to decide whether there exists a suitable catalyst with dimension
not smaller than $k$ since any potential catalyst should have a
dimensionality not smaller  than $k$.

To conclude this section,  we consider another conjecture made by
Nielsen. More specifically, suppose  a transformation
$x\rightarrow y$ can be catalyzed by a catalyst state $c$, where
$x$  and $y\in V^n$. One may naturally hope that the dimension of
$c$ is not too large, for example, bounded from top  by $n$ or
$n^2$. This conjecture, first addressed by Nielsen \cite{MAJ}, was
proved to be false in \cite{DK01}. More precisely,  Daftuar $et\
al$ showed that if $S(y)\subsetneqq T(y)$ then $T_k(y)\neq T(y)$
for any $k\geq 1$. In other words, if catalysis is useful for $y$,
then the dimension of catalyst is not bounded, and thus yields a
negative answer for Nielsen's conjecture. Since the proof
presented in \cite{DK01} is not a constructive one, we still do
not know that what kinds of states $y$ have such strange
properties. Theorem \ref{kmelocc} characterizes precisely the
lower bound of the dimension of catalyst $c$ for any state $y$,
also the lower bound of the number of copies of any multiple-copy
entanglement transformations to $y$. We give an concrete  example
as follows:
\begin{example}\label{example1}\upshape
Take  $y^{k}=(1,\alpha,\alpha^k,\beta)/C,$ where $k>1$,
$0<\alpha<1$, $0\leq \beta< \alpha^{k+2}$ and
$C=1+\alpha+\alpha^k+\beta.$

By Theorem \ref{kmelocc}, we have $T_k(y^{k})=M_k(y^{k})=S(y^{k})$
but $S(y^k)\subsetneqq T_{k+1}(y^k){\rm \ and\ } S(y^k)\subsetneqq
M_{k+1}(y^k)$ for any $k>1$. Such a state $y^k$ has a very strange
property: although it can be catalyzed by some catalysts, any
state with dimension less than $k$ cannot serve as catalyst for
it.

For example, if we take $k=16$, the state $y^{16}$ is
$4$-dimensional, but it has no quantum catalyst $c$ with dimension
not more than $16$. Suppose that $x$ can be catalyzed into $y$.
Then the catalyst state $c$ should have a dimensionality  at least
$17$.  We also have that any multiple-copy entanglement
transformations with the number of copies less than $17$ have no
advantage.
\end{example}

\section{Some lemmas}

From this section on, we are going to give detailed proofs of the
results stated in Section II. The purpose of this section is to
collect some lemmas needed in the proofs. For the sake of
convenience, we introduce some notations of set operations. Let
$A$ and $B$ be two sets of finite dimensional vectors. Then
$A\oplus B$ denotes the set of all vectors of the form $a\oplus b$
with $a\in A$ and $b\in B$, i.e., $A\oplus B=\{a\oplus b: a\in
A{\rm\ and\ }b\in B\}$. Similarly, $A\otimes B=\{a\otimes b: a\in
A {\rm\ and \ } b\in B\}$. If $c$ is a vector, then $A\otimes c$
is just a convenient form of $A\otimes \{c\}$.  Note that $a\oplus
b $ denotes the direct sum of vectors, that is, the vector
concatenating $a$ and $b$, or $(a,b)$.

First, we recall some  simple properties of  majorization from
\cite{MO79}.
\begin{lemma}\label{directum}\upshape For any $y$ and $y'$, we have $S(y)\oplus S(y')\subseteq S(y\oplus y')$
and $S(y)\otimes S(y')\subseteq S(y\otimes y')$. That is, $x\prec
y$ and $x'\prec y'$ imply $x\oplus x'\prec y\oplus y'$ and
$x\otimes x'\prec y\otimes y'$.
\end{lemma}

The major difficulties in studying the structure of entanglement
catalysis and multiple-copy entanglement transformation are the
lack of suitable mathematical tools to deal with  tensor product
and marjorization relation. In what follows, we try to provide
some useful tools to overcome these difficulties. They are mainly
about the strict majorization relation under  direct sum and
tensor product.

For a subset $A\subseteq V^n$, the set of all interior points of
$A$ is denoted by $A^{o}$. It is easy to check that $x\in
S^{o}(y)$ if and only if $x\lhd y$. We also note that
$S^o(y)=\emptyset$ only occurs in the case of $y_n=y_1$, which
means that $y$ is a uniform vector. Without clearly stating, we
always assume that $S^o(y)\neq \emptyset$. Extreme components of a
vector are used frequently. For simplicity, we denote by ${\rm
max\ }x$ and ${\rm min\ }x$ the maximal and the minimal components
of $x$, respectively.

The following lemma is crucial in this paper. It says, to keep the
direct sum of the interiors of $S(y')$ and $S(y'')$ still in the
interior of $S(y'\oplus y'')$, $y'$ should suitably overlap with
$y''$, and vice versa.

\begin{lemma}\label{interiorpointdirectsum}\upshape
For any $y$ and $y'$, we have
\begin{equation}\label{interiorsumequi}
S^{o}(y)\oplus S^{o}(y')\subseteq S^{o}(y\oplus y')
\Leftrightarrow {\rm max\ }y>{\rm min\ }y'{\rm \ and\ } {\rm max\
}y'>{\rm min\ }y.
\end{equation}
That is, if  $x\lhd y$ and $x'\lhd y'$, then $x\oplus x'\lhd
y\oplus y'$ if and only if the conditions in right-hand side of
Eq.(\ref{interiorsumequi}) hold.
\end{lemma}

A careful observation of the proof of Lemma
\ref{interiorpointdirectsum} shows that the sets $S^o(y)\oplus
S^o(y')$ and $S^o(y\oplus y')$ satisfy an interesting property: if
there exists $z$ such that $z\in S^o(y)\oplus S^o(y')$ and $z\in
S^o(y\oplus y')$, then for any $\bar{z}\in S^o(y)\oplus S^o(y')$
it holds that $\bar{z}\in S^o(y\oplus y')$. Since we will use this
property considerably latter, we formalize it as the following:
\begin{definition}\label{linearproperty}
We say that two nonempty sets $A$ and $B$ satisfy linearity
property (LP), if $A\cap B\neq \emptyset$ implies $A\subseteq B$.
\end{definition}

Before stating a corollary of Lemma \ref{interiorpointdirectsum},
we introduce a useful notation. We use $x^{\oplus k}$ to denote
$k$ times direct sum of $x$ itself. That is, $x^{\oplus
k}=\underbrace{x\oplus x\oplus \cdots \oplus x}_{k{\rm \ times\
}}. $ Similarly, for a set $A$, $A^{\oplus k}=\underbrace{A\oplus
A\oplus \cdots \oplus A}_{k{\rm\ times\ }}.$

Now a direct consequence of Lemma \ref{interiorpointdirectsum} is
as follows:
\begin{corollary}\label{sumrepeat}\upshape
For any $y$ and  $k\geq 1$, $(S^o(y))^{\oplus k}\subseteq
S^o(y^{\oplus k})$. Specially, $x\lhd y\Rightarrow x^{\oplus
k}\lhd  y^{\oplus k}.$
\end{corollary}

Combining Lemma \ref{interiorpointdirectsum} with Corollary
\ref{sumrepeat} we obtain the following necessary and sufficient
condition for determining  whether a given $x$ is in the interior
of $S(y)$.
\begin{corollary}\label{sumrepeatset}\upshape
Suppose  $\{(y^i)^{\oplus k_i}:1\leq i\leq m\}$ is a set of
vectors and $x^i\lhd y^i$ for any $1\leq i\leq m$. Denote
$x=\oplus_{i=1}^m (x^i)^{\oplus k_i}$,
$y=\oplus_{i=1}^m(y^i)^{\oplus k_i}$. Then the following are
equivalent:

(1) $x\lhd y$. Or by LP, $\oplus_{i=1}^m (S^o(y^i))^{\oplus
k_i}\subseteq S^o(\oplus_{i=1}^m (y^i)^{\oplus k_i})$.

(2) There exist $1\leq j_1<j_2<\cdots<j_t\leq m$ such that  (i)
${\rm max\ }y^{j_1}={\rm max}\{{\rm max\ }y^i:1\leq i\leq m\},
{\rm min\ }y^{j_t}={\rm min}\{{\rm min\ }y^i:1\leq i\leq m\}$, and
(ii) ${\rm min\ }y^{j_s}<{\rm max\ }y^{j_s+1}{\rm \ for \ each\
}1\leq s\leq t-1$.
\end{corollary}

Intuitively, the sequence of $y^{j_1},\cdots, y^{j_t}$ is called
an overlapping sequence of the set $\{y^{\oplus k_i}:1\leq i\leq
m\}$.

By using Corollary \ref{sumrepeatset}, we have the following
powerful lemma dealing with tensor product.

\begin{lemma}\label{lginterior}\upshape
For any $y$ and $c$,  $S^{o}(y)\otimes c\subseteq S^{o}(y\otimes
c)$ if and only if $l_u(c)>g_u(y).$
\end{lemma}

\textit{Proof.} Let $y\in V^n$ and $c\in V^k$. Take $x\lhd y$. We
consider the set of vectors $\{c_iy: i=1,\cdots, k\}$. If
$l_u(c)>g_u(y)$ then by the definitions of uniformity indices, we
have $c_iy_n<c_{i+1}y_1$ for all $1\leq i<k$ , which can be
restated as ${\rm min\ } c_iy<{\rm max\ }c_{i+1}y$. Applying
Corollary \ref{sumrepeatset} yields
$$\oplus_{1\leq i\leq k}c_ix\lhd \oplus_{1\leq i\leq k}c_iy,$$ which is the same as $x\otimes c\lhd y\otimes c$.

Conversely, if $l_u(c)\leq g_u(y)$, then there should exist an
$1\leq i_0\leq k-1$ such that $c_{i_0+1}/c_{i_0}\leq y_n/y_1$, or
${\rm min\ } c_{i_0}y\geq {\rm max\ }c_{i_0+1}y.$ Thus
$$e_{i_0n}(y\otimes
c)=\sum_{i=1}^{i_0}e_n(c_iy)=\sum_{i=1}^{i_0}e_n(c_ix)\leq
e_{i_0n}(x\otimes c),$$ which contradicts $x\otimes c\lhd
y\otimes c$. Thus we complete the proof of the lemma.\hfill $\square$\\

With the aid of Lemma \ref{lginterior}, we can show that the
interior of $S(y)$ is closed under tensor product, as the
following lemma indicates:
\begin{lemma}\label{interiorpointproduct}\upshape
For any  $y$ and $y'$,  $ S^{o}(y)\otimes S^o(y')\subseteq
S^o(y\otimes y').$
\end{lemma}

\textit{Proof.} Take $x\lhd y$ and $x'\lhd y'$. Then by Lemma
\ref{directum}, it follows that
\begin{equation}\label{eq8}
x\otimes x'\prec x\otimes y'\prec y\otimes y'.
\end{equation}
If one of the inequalities  in Eq.(\ref{eq8}) is strict, then we
have done. Otherwise, by Lemma \ref{lginterior}, the first
inequality in Eq.(\ref{eq8}) is not strictly implies $l_u(x)\leq
g_u(y').$ Similarly, the second inequality in Eq.(\ref{eq8}) is
not strictly implies $l_u(y')\leq g_u(y).$ By Eq.(\ref{lugu}), we
have $g_u(x)\leq l_u(x)$ and $g_u(y')\leq l_u(y')$. Hence we have
$g_u(x)\leq g_u(y)$. However, $x\lhd y$ implies $x_1<y_1$ and
$x_n>y_n$, which yield
$g_u(x)>g_u(y)$, a contradiction.\hfill $\square$\\

An immediate consequence of Lemma \ref{interiorpointproduct} is
the following:
\begin{corollary}\label{corointerior}
For positive integers $k$, $p$, and $q$, $(S^o(y))^{\otimes
k}\subseteq S^o(y^{\otimes k})$ and $(S^o(y))^{\otimes p}\otimes
(S^o(y'))^{\otimes q}\subseteq S^o(y^{\otimes p}\otimes
y'^{\otimes q}).$
\end{corollary}

Properties of strict super majorization are much more simpler than
that of strict majorization, and we list some of them  as follows:

\begin{lemma}\label{strictsuper}
For any non-negative vectors $x$, $y$, $x'$ and $y'$, we have

1) if $x\lhd^w y$ and $x'\lhd^w y'$, then $x\oplus x'\lhd^w
y\oplus y'$.

2) if $x\lhd^w y$ and $x'\lhd y'$, then $x\oplus x'\lhd^w y\oplus
y'$ if and only if ${\rm max\ }y'>{\rm min\ }y$.

3) if $x\lhd^w y$ and $x'\prec^w y'$, then $x\otimes  x'\lhd^w
y\otimes y'$. Here we assume ${\rm min\ }x'>0$.
\end{lemma}

Usually, we use the following generalized version of $2)$ of the
above lemma:
\begin{corollary}\label{repeatsupermaj}
Suppose that $x=(x^1,\cdots, x^m)$ and $y=(y^1,\cdots, y^m)$
satisfy $x^i\lhd y^i$ for each $1\leq i\leq m$, and  $x^0\lhd^w
y^0$. Then $x^0\oplus x\lhd^w y^0\oplus y$ if and only if ${\rm
max\ }y^m>{\rm min\ }y^0$.
\end{corollary}

\section{Proof of Theorem \ref{catalystuseful}}
The main aim of this section is to prove Theorem
\ref{catalystuseful}. First, we present some necessary
preliminaries. Especially, we give  a characterization of
$T^o(y,c)$, and then  a necessary and sufficient condition for
$S(y)\subsetneqq T(y,c)$, which also leads to an efficient
algorithm to solve the problem that whether $c$ is useful for $y$.
In addition to that, two auxiliary lemmas are also introduced.
With those we can finish the proof of Theorem
\ref{catalystuseful}.

A decomposition of a catalyst state is a useful mathematical tool
in this section. Formally, we have the following:
\begin{definition}\label{reldecom}\upshape
We say $[c]_\alpha=\{c^1,\cdots,c^m\}$ is a decomposition of $c$
according to $\alpha$ with $0\leq \alpha<1$, if

1)  $c=c^1\oplus \cdots \oplus c^m$;

2) $l_u(c^i)>\alpha$ for all $1\leq i\leq  m$;

3) ${\rm max\ }c^{i+1}/{\rm min\ }c^i \leq  \alpha$ for all $1\leq
i\leq m-1$.
\end{definition}
Obviously, for any $\alpha\in[0,1)$ and $c\in V^k$, the
decomposition $[c]_\alpha$ exists uniquely. Given $c$ and
$\alpha$, it is only a simple calculation to determine
$[c]_\alpha$. We shall see that such a decomposition plays an
important role in the study of entanglement catalysis.

Two useful lemmas are needed to present a simple characterization
of $T^o(y,c)$. The first lemma shows the importance of the
decomposition of $c$ according to $g_u(y)$:
\begin{lemma}\label{tyanddecom}\upshape
If $[c]_{g_u(y)}=\{c^1,\cdots, c^m\}$,  then $T(y,c)=\cap_{i=1}^m
T(y,c^i)$.
\end{lemma}

\textit{Proof.} We only need to prove that $x\otimes c\prec
y\otimes c$ if and only if $x\otimes c^i\prec y\otimes c^i$ for
each  $1\leq i\leq m$. Sufficiency part follows immediately from
Lemma \ref{directum}. Now we prove the necessity part. By the
definition of $[c]_{g_u(y)}$,  we have $ g_u(y)\geq {\rm max\
}c^{i+1}/{\rm min\ }c^i$, that is, ${\rm max\ }y\otimes
c^{i+1}\leq {\rm min\ }y\otimes c^i$ for all $1\leq i<m$, which
follows that
$$(y\otimes c)^{\downarrow}=((y\otimes c^1)^{\downarrow},\cdots, (y\otimes c^m)^{\downarrow}).$$

Noticing  $x\otimes c\prec y\otimes c$ implies  $x_1\leq y_1$ and
$x_n\geq y_n$, we have $g_u(x)\geq g_u(y)$, thus
$$(x\otimes c)^{\downarrow}=((x\otimes c^1)^{\downarrow},\cdots, (x\otimes c^m)^{\downarrow}).$$
Hence,  majorization relation $x\otimes c\prec y\otimes c$ splits
into $m$ sub-majorizations: $x\otimes c^i\prec y\otimes c^i$,
$1\leq i\leq m$. \hfill $\square$\\

By virtue of Lemma \ref{tyanddecom},  we only need to focus on the
case that the catalyst $c$ and the target $y$ satisfy
$l_u(c)>g_u(y)$. In this special case,  the following lemma  shows
that $x\in T^o(y,c)$ is just equivalent to $x\otimes c \lhd
y\otimes c$.
\begin{lemma}\label{tyinterior}\upshape
If $l_u(c)>g_u(y)$, then $T^o(y,c)=\{x\in V^n:x\otimes c\lhd
y\otimes c\}$.
\end{lemma}

\textit{Proof.} To make the paper  more readable, the lengthy
proof is put into
Appendix B.\hfill $\square$\\

Combining the above two lemmas leads us to the following simple
characterization of $T^o(y,c)$:
\begin{theorem}\label{tycinterior}\upshape
Let $[c]_{g_u(y)}=\{c^1,\cdots, c^m\}$. Then
$T^o(y,c)=\{x:x\otimes c^i\lhd y\otimes c^i {\rm\ for\ any\ }1\leq
i\leq m\}$.
\end{theorem}

\textit{Proof.} We only need to prove that if $x\in T^o(y,c)$,
then $x\otimes c^i\lhd y\otimes c^i$ for all $1\leq i\leq m$. In
fact, by Lemma \ref{tyanddecom}, $x\in T^o(y,c)$ implies $x\in
T^o(y,c^i)$. Since $l_u(c^i)> g_u(y)$, it follows that $x\otimes
c^i\lhd y\otimes c^i$ by Lemma \ref{tyinterior}. \hfill $\square$\\

To present the condition of the inequality $S(y)\subsetneqq
T(y,c)$ compactly, we introduce a special probability vector
$y(d)$ for each $1<d<n-1$. Formally, $y(d)$ is defined as
$y_i(d)=e_d(y)/d$ if $1\leq i\leq d$, and
$y_i(d)=(e_n(y)-e_d(y))/(n-d)$ if $d+1\leq i\leq n$.  Now we have
the following:
\begin{theorem}\label{characterizationforcatalystuseful}\upshape
For any $y\in V^n$,  $S(y)\subsetneqq T(y,c)$ if and only if there
exists $1<d<n-1$ such that $y(d)\otimes c'\lhd y\otimes c'$ for
any $c'\in [c]_{g_u(y)}$.
\end{theorem}

\textit{Proof .} We first deal with the sufficiency part. Since
$y(d)\otimes c'\lhd y\otimes c'$ for each $c'\in [c]_{g_u(y)}$, we
have $y(d)\in T^o(y,c)$ by Theorem \ref{tycinterior}. On the other
hand, $y(d)$ is a boundary point of $S(y)$ as $e_d(y(d))=e_d(y)$.
These and the fact that $S(y)\subseteq T(y,c)$ yield
$S(y)\subsetneqq T(y,c)$.

Conversely, assume $S(y)\subsetneqq T(y,c)$.  It is easy to verify
that both $S(y)$ and $T(y,c)$ are compact subsets of $V^n$. Thus
$S(y)\subsetneqq T(y,c)$ implies $S^o(y)\subsetneqq T^o(y,c)$. In
other words, there should exist some $x$ such that $x$ is a
boundary point of $S(y)$ while $x$ is in the interior of $T(y,c)$.
$x$ is a boundary point of $S(y)$ implies that there exists some
$1<d<n-1$ such that $e_d(x)=e_d(y)$. By the definition of $y(d)$,
one can easily see $y(d)\prec x$, which together with $x\in
T^o(y,c)$ yields $y(d)\in T^o(y,c)$. By Theorem \ref{tycinterior},
we have $y(d)\otimes c'\lhd y\otimes c'$ for each $c'\in
[c]_{g_u(y)}$.
\hfill $\square$\\

Theorem \ref{characterizationforcatalystuseful} provides a
complete characterization of $S(y)\subsetneqq T(y,c)$. It also
tells us how to simply determine whether a catalyst state $c$ is
useful for a  target state $y$. More precisely, for a given
$k$-dimensional probability vector $c$ and $n$-dimensional
probability vector $y$, to check whether $S(y)\subsetneqq T(y,c)$,
we only need to verify that whether $y(d)\otimes c'\lhd y\otimes
c'$ for each $c'\in [c]_{g_u(y)}$ and some $1<d<n-1$. By a simple
calculation, it is easy to see that the total time complexity of
this problem is $O(n^2k\log(nk))$. Thus this problem can be
efficiently solved.

For $1<d<n-1$, we introduce a useful subset of $S(y)$.
Specifically, we define $K_d(y)=S^o(y')\oplus S^o(y'')$, where
$y'=(y_1,\cdots, y_d)$ and $y''=(y_{d+1},\cdots, y_n)$. Obviously,
$K_d(y)=\emptyset$ if and only if $y_1=y_d$ or $y_{d+1}=y_n$. In
what follows, we always assume $K_d(y)\neq \emptyset$ except we
clearly say that it is not the case. For any $x\in K_d(y)$, we
have $y(d)\prec x$. From this point of view, $y(d)$ can be treated
as a center of $K_d(y)$.

It is easy to check that $K_d(y)\otimes c$ and $S^o(y\otimes c)$
satisfy LP. So we have the following interesting consequence of
Theorem \ref{characterizationforcatalystuseful}:
\begin{corollary}\label{alternative}\upshape
For any $y\in V^n$, $S(y)\subsetneqq T(y,c)$ if and only if there
exists $1<d<n-1$ such that $K_d(y)\otimes c'\subseteq S^o(y\otimes
c')$ for any $c'\in [c]_{g_u(y)}$.
\end{corollary}

To give a proof of Theorem \ref{catalystuseful}, the following two
simple lemmas are needed.

The first lemma provides a simple sufficient condition for
$K_d(y)\otimes c\subseteq S^o(y\otimes c)$.
\begin{lemma}\label{sufficient}\upshape
For $y\in V^n$ and $1<d<n-1$, if a catalyst  $c$ satisfies
\begin{equation}\label{newcond10}
l_u(c)>{\rm max}\{\frac{y_d}{y_1}, \frac{y_n}{y_{d+1}}\} {\rm\
and\ } g_u(c)<\frac{y_{d+1}}{y_d},
\end{equation}
then $K_d(y)\otimes c \subseteq S^o(y\otimes c)$.
\end{lemma}
\textit{Proof.} For any $x\in K_d(y)$, we will show that $x\otimes
c\in S^o(y\otimes c)$. For this purpose, let us decompose
$x=(x',x'')\ {\rm\ and\ } \ y=(y',y''),$ where $x'$ is formed by
the largest $d$ components of $x$, $x''$ is the rest part, and
$y'$ and $y''$ are defined similarly. It is obvious that $x'\lhd
y'$ and $x''\lhd y''$. Noticing that $l_u(c)>{\rm
max}\{g_u(y'),g_u(y'')\}$, we have $x'\otimes c\lhd y'\otimes c$
and $x''\otimes c\lhd y''\otimes c$ by Lemma \ref{lginterior}.
Furthermore,
$$g_u(c)<\frac{y_{d+1}}{y_{d}}$$ is equivalent to
$${\rm max\ }y'\otimes c>{\rm min\ }y''\otimes c$$ and
$${\rm max\ }y''\otimes c>{\rm min\ }y'\otimes c.$$
These facts imply that $x'\otimes c \oplus x''\otimes c\lhd
y'\otimes c\oplus y''\otimes c$ according to Lemma
\ref{interiorpointdirectsum}. Equivalently, we have $x\otimes
c\lhd y\otimes c$, which completes  the proof of $x\otimes c\in
S^{o}(y\otimes c)$. \hfill $\square$\\

The second lemma provides a necessary condition for $K_d(y)\otimes
c \subseteq S^o(y\otimes c)$.
\begin{lemma}\label{necessary}\upshape
If $K_d(y)\otimes c \subseteq S^o(y\otimes c)$, then there exists
a segment of $c$, namely $c'$,  such that
\begin{equation}\label{newcond}
l_u(c')>{\rm max}\{\frac{y_d}{y_1}, \frac{y_n}{y_{d+1}}\} {\rm\
and\ } g_u(c')<\frac{y_{d+1}}{y_d}.
\end{equation}
\end{lemma}

\textit{Proof.} Take $x\in K_d(y)$. Then $x\otimes c\lhd y\otimes
c$. We will explicitly construct $c'$ that is a segment of $c$
satisfying Eq.(\ref{newcond}).  Similar to Lemma \ref{sufficient},
we decompose $x=(x',x'')$ and $y=(y',y'')$. By the definition of
$K_d(y)$, we have $x'\lhd y'$ and $x''\lhd y''$.

Denote $$\alpha={\rm max}\{\frac{y_d}{y_1},
\frac{y_n}{y_{d+1}}\},$$ and  decompose $c$ according to $\alpha$
into  $[c]_{\alpha}=\{c^1,\cdots, c^m\}$. Then the vector $c'$
satisfying Eq.(\ref{newcond}) can be constructed as follows: If
$g_u(y')\geq g_u(y'')$, then $c'=c^1$; otherwise $c'=c^m$. We only
prove the case of $g_u(y')\geq g_u(y'')$. In this case, we have
$\alpha=g_u(y')$. Since $c^1\in [c]_\alpha$, it satisfies
$l_u(c^1)>\alpha$.  The only left thing is to prove
$g_u(c^1)<{y_{d+1}}/{y_d}$. By contradiction, suppose that
$g_u(c^1)\geq {y_{d+1}}/{y_d}$. This relation  can be restated as
\begin{equation}\label{eq3}
{\rm min\ }y'\otimes c^1\geq {\rm max\ }y''\otimes c^1.
\end{equation}
By the definition of $[c]_{\alpha}$, we also have
\begin{equation}\label{eq4}
{\rm min\ }y'\otimes c^1\geq {\rm max\ }y'\otimes c^2.
\end{equation}
Therefore, if we take $l={\rm dim}(y'\otimes c^1)$, then by
Eqs.(\ref{eq3}) and (\ref{eq4}), the $l$ largest components of
$y\otimes c$ are just those of $y'\otimes c^1$, which yields
$$e_l(y\otimes c)=e_l(y'\otimes c^1)=e_l(x'\otimes c^1)\leq e_l(x\otimes c),$$
a contradiction with $x\otimes c\lhd y\otimes c$.

The case of $g_u(y')<g_u(y'')$ can be proved similarly
by considering the term $y''\otimes c^m$. With that we complete the proof of Lemma \ref{necessary}.\hfill $\square$\\

Now the proof of Theorem \ref{catalystuseful} follows immediately.

\textit{Proof of Theorem \ref{catalystuseful}.}  If
Eq.(\ref{usecata}) holds, then by Lemma \ref{sufficient}, we have
$K_d(y)\otimes c\subseteq S^o(y\otimes c)$, which follows
$S(y)\subsetneqq T(y,c)$ by Corollary \ref{alternative}.

Conversely, if $S(y)\subsetneqq T(y,c)$, then by Corollary
\ref{alternative}, we have $K_d(y)\otimes z\subseteq S^o(y\otimes
z)$ for some $z\in [c]_{g_u(y)}$ and $1<d<n-1$. Moreover,
application of Lemma \ref{necessary} indicates the existence of
$c'$ satisfying Eq.(\ref{newcond}), or equivalently,
Eq.(\ref{usecata}), where $c'$ is a segment of $z$, and also a
segment of $c$. With that we complete the proof of the Theorem
\ref{catalystuseful}.\hfill $\square$\\

\section{Proof of Theorem \ref{kmelocc}}

In this section, we will give a proof of Theorem \ref{kmelocc}.
First,  some necessary preliminaries are presented. Especially, we
obtain characterizations of $T_k^o(y)$ and $M_k^o(y)$,
respectively. Then we  propose a proof of Theorem \ref{kmelocc}.

\begin{theorem}\label{tkystructure}\upshape
For any $y$, $T_k^o(y)=\{x: x\otimes c\lhd  y\otimes c {\rm\ for\
some\ } c\in V^{k'},k'\leq k\}$. Similarly, $T^o(y)=\{x: x\otimes
c\lhd y\otimes c {\rm\ for\ some\ } c\}$.
\end{theorem}

\textit{Proof.} The part that $x\otimes c\lhd y\otimes c$ implies
$x\in T_k^o(y)$ is obvious. Conversely, suppose that $x$ is an
interior point of $T_k(y)$. Let us define
$\bar{x}=(x_1+\epsilon,\cdots, x_n-\epsilon)$. Since $x\in
T_k^o(y)$,  we have $\bar{x}\in T_k(y)$ for a sufficiently small
positive real $\epsilon$, or equivalently, $\bar{x}\otimes c\prec
y\otimes c$ for some $c\in V^k$. On the other hand, by the
construction, it is easy to check that $x\lhd \bar{x}$. Thus $x\in
T^o(y,c)$. By Theorem \ref{tycinterior}, we have $x\otimes c'\lhd
y\otimes c'$ for some $c'\in V^{k'}$, where $k'\leq k$.

The case of $T(y)$ can be similarly proved. \hfill $\square$\\

\begin{theorem}\label{mkystructure}\upshape
For any  $y\in V^n$,  $M_k^o(y)=\{x: x^{\otimes k }\lhd
y^{\otimes k}\}$. Similarly, $M^o(y)=\{x: x^{\otimes k }\lhd
y^{\otimes k}{\rm\ for\ some\ } k\geq 1\}$.
\end{theorem}

\textit{Proof.} The part that $x^{\otimes k}\lhd y^{\otimes k}$
implies $x\in M_k^o(y)$ is obvious. Conversely, suppose that $x$
is an interior point of $M_k(y)$. Let us define
$\bar{x}=(x_1+\epsilon,\cdots, x_n-\epsilon)$. Since $x\in
M_k^o(y)$, we have $\bar{x}\in M_k(y)$ for a sufficiently small
positive real $\epsilon$, or equivalently, $\bar{x}^{\otimes
k}\prec y^{\otimes k}$. On the other hand, by the construction, we
have $x\lhd \bar{x}$. Applying of Corollary \ref{corointerior}
yields $x^{\otimes k}\lhd \bar{x}^{\otimes k}$. Hence $x^{\otimes
k}\lhd y^{\otimes k}$.

The case of $M(y)$ can be similarly proved. \hfill $\square$\\

The following lemma is a powerful tool in the study of the
mathematical structure of MLOCC.
\begin{lemma}\label{uniqueinterior}\upshape
For any $x\in K_d(y)$, we have
\begin{equation}\label{2interiorcondition}
x^{\otimes k}\lhd y^{\otimes k} \Leftrightarrow
 y_d^k<y_1^{k-1}y_{d+1} {\rm\ and\ }
                y_{d+1}^k>y_dy _n^{k-1}.
\end{equation}
\end{lemma}

\textit{Proof.} We put the lengthy proof of this lemma
in Appendix C. \hfill $\square$\\

Now we present a very interesting result. In fact, we are able to
completely characterize  the condition of $K_d(y)\subseteq
T_k^o(y)$ and  that of $K_d(y)\subseteq M_k^o(y)$. To one's
surprise, these two conditions are exactly the same. As we will
see soon, the equivalence  of $K_d(y)\subseteq T_k^o(y)$ and
$K_d(y)\subseteq M_k^o(y)$ directly leads us to the proof of
Theorem \ref{kmelocc}:

\begin{theorem}\label{minterior}\upshape
For any $y\in V^n$ and $1<d<n-1$, the following are equivalent:

1) $K_d(y)\subseteq T_k^o(y)$.

2) $K_d(y)\subseteq M_k^o(y)$.

3) $y_d^{k}<y_1^{k-1}y_{d+1}$ and $y_{d+1}^k>y_n^{k-1}y_d$.

\end{theorem}

\textit{Proof.} We first prove the equivalence of 1) and 3).
Suppose that 3) holds. Then we can choose
$c=(1,\alpha,\cdots,\alpha^{k-1})$ with $0<\alpha<1$ such that
$${\rm max\ }\{(\frac{y_d}{y_1})^{k-1},
(\frac{y_n}{y_{d+1}})^{k-1}\}<\alpha^{k-1}<\frac{y_{d+1}}{y_d}.$$
A routine calculation shows that $l_u(c)=\alpha{\rm \ and  \ }
g_u(c)=\alpha^{k-1}.$  By Lemma \ref{sufficient}, we have
$K_d(y)\subseteq T^o(y,c)\subseteq T_k^o(y).$

Conversely, by Theorem \ref{tkystructure}, $K_d(y)\subseteq
T_k^o(y)$ implies that there exists $c\in V^{k'}$ such that
$y(d)\otimes c\lhd  y\otimes c,$ and then by LP we have
$K_d(y)\otimes c\subseteq S^o(y\otimes c)$.  According to Lemma
\ref{necessary}, we declare that there exists $z\in V^{k''}$
satisfying Eq.(\ref{newcond}), where $z$ is a segment of $c$.
Noticing that $ l^{k''-1}_u(z)\leq g_u(z)$ and $k''\leq k'\leq k$,
we have $l^{k-1}_u(z)\leq g_u(z)$. This fact together with
Eq.(\ref{newcond}) shows that $$y_d^k< y_1^{k-1}y_{d+1} {\rm \
and\ }y_{d+1}^k> y_dy_n^{k-1},$$ which is exactly the same as 3).

By Theorem \ref{mkystructure}, $K_d(y)\subseteq M_k^o(y)$ is
equivalent to that for any $x\in K_d(y)$, $x^{\otimes k}\lhd
y^{\otimes k}$. Therefore, the equivalence of 1) and 3) follows
from the following fact: for any $x\in K_d(y)$, $x^{\otimes k}\lhd
y^{\otimes k}$ if and only if $y_d^{k}<y_1^{k-1}y_{d+1}$ and
$y_{d+1}^k>y_n^{k-1}y_d$. Obviously, this fact  is guaranteed by Lemma \ref{uniqueinterior}. \hfill $\square$\\

We are now in the position to present the proof of  Theorem
\ref{kmelocc}.

\textit{Proof of Theorem \ref{kmelocc}.} Theorem \ref{kmelocc} is
essentially implied by Theorem \ref{minterior}. To see this, we
only need to prove that $S(y)\subsetneqq T_k(y)$ and
$S(y)\subsetneqq M_k(y)$ are equivalent to $K_d(y)\subseteq
T_k^o(y)$ and $K_d(y)\subseteq M_k^o(y)$ for some $1<d<n-1$,
respectively. The proofs of these two cases are similar.  We only
outline the proof for the case of $M_k(y)$ here.  The part that
$K_d(y)\subseteq M_k^o(y)$ implies  $S(y)\subsetneqq M_k(y)$ is
obvious. Conversely, assume that $S(y)\subsetneqq M_k(y)$.
Noticing that $M_k(y)$ is a compact subset of $V^n$, we can find
some $1<d<n-1$ such that $y(d)\in M_k^o(y)$ by a similar argument
used in the proof of Theorem
\ref{characterizationforcatalystuseful}. Now $K_d(y)\subseteq
M_k^o(y)$ follows from LP. \hfill $\square$\\

\section{Proofs of Theorems \ref{characterizationforcatalystusefulprob} and \ref{kemloccprob}}

In this section, we  mainly prove  Theorems
\ref{characterizationforcatalystusefulprob} and \ref{kemloccprob}.
First, the  physical meaning of the interior points of
probabilistic entanglement transformations is given. Then we
complete the proof of Theorem
\ref{characterizationforcatalystusefulprob}. Finally, the proof of
Theorem \ref{kemloccprob} is given. In this section, we always
assume that $\lambda\in (0,1)$.

As we have mentioned, the interior point of $S^\lambda(y)$ has a
very clear physical meaning. That is, $x$ is an interior point of
$S^\lambda(y)$ if and only if the maximal conversion probability
from $x$ to $y$ is strictly larger than $\lambda$. Equivalently,
we have $x\in (S^\lambda(y))^o$ if and only if $x\lhd^w \lambda
y$. An interesting question is thus to ask whether this property
also holds in the presence of catalysts or in multiple-copy
scheme. The following result gives a positive answer to this
question in the case of probabilistic ELOCC:
\begin{theorem}\label{eloccprobinterior}\upshape
For any $y\in V^n$, it holds that $(T^\lambda(y,c))^o=\{x\in
V^n:x\otimes c\lhd^w \lambda y\otimes c \}$. Similarly, we have
$(T_k^\lambda(y))^o=\{x\in V^n: x\otimes c\lhd^w \lambda y\otimes
c {\rm\ for\ some\ }c\in V^k\}$ and $(T^\lambda(y))^o=\{x\in V^n:
x\otimes c\lhd^w \lambda y\otimes c {\rm\ for\ some\ }c\}$.
\end{theorem}

\textit{Proof.} The proof is similar to Lemma \ref{tyinterior}. We
omit the details here.\hfill $\square$\\

We can  prove a corresponding result of the above theorem in the
case of probabilistic  MLOCC:
\begin{theorem}\label{mloccprobinterior}
For any $y\in V^n$, we have $(M_k^\lambda(y))^o=\{x\in V^n:
x^{\otimes k}\lhd^w \lambda^k y^{\otimes k}\}$. Similarly,
$(M^\lambda(y))^o=\{x\in V^n: x^{\otimes k}\lhd^w \lambda^k
y^{\otimes k} {\rm\ for\ some\ }k\geq 1\}$.
\end{theorem}

\textit{Proof.} The proof is almost the same as that of Theorem \ref{mkystructure}. So we omit the details here.\hfill $\square$\\

We can extend $K_d(y)$ in a probabilistic manner. Formally, define
$K_d^{\lambda}(y)=\{x\in S^{\lambda}(y): E_l(x)=\lambda
E_l(y){\rm\ iff\ } l=n-d\}$, where $0<d<n-1$.

The proof of Theorem \ref{characterizationforcatalystusefulprob}
goes  as follows:

\textit{Proof of Theorem
\ref{characterizationforcatalystusefulprob}.} Similar to the
deterministic cases, it is easy to show that
$S^\lambda(y)\subsetneqq T^\lambda(y,c)$ if and only if there
exists $0<d<n-1$ such that $K_d^\lambda(y)\otimes c\subseteq
(S^\lambda(y\otimes c))^o$. Hence, to complete the proof, we need
only to show that $K_d^\lambda(y)\otimes c\subseteq
(S^\lambda(y\otimes c))^o$ if and only if
\begin{equation}\label{probusecata2}
l_u(c^m)>\frac{y_n}{y_{d+1}} {\rm\ and\
}g_u(c^m)<\frac{y_{d+1}}{y_d},
\end{equation}
where $c^m$ is  the last element of $[c]_{g_u(y'')}=\{c^1,\cdots,
c^m\}$.

Take $x\in K_d^\lambda(y)$, and decompose $x=(x',x'')$ and
$y=(y',y'')$. It is obvious that
$$x'\lhd^w \lambda
y' {\rm\ and\ } x''\lhd \lambda y''.$$

By Lemma \ref{strictsuper}, we have $x'\otimes c\lhd^w \lambda
y'\otimes c$. From the definition  of $[c]_{g_u(y'')}$, we know
$x''\otimes c^i\lhd \lambda y''\otimes c^i$ for any $1\leq i\leq
m$, and furthermore,
$$(y''\otimes c)^\downarrow=((y''\otimes c^1)^\downarrow,\cdots, (y''\otimes c^m)^\downarrow)$$
and
$$(x''\otimes c)^\downarrow=((x''\otimes c^1)^\downarrow,\cdots, (x''\otimes c^m)^\downarrow).$$

According to Corollary \ref{repeatsupermaj}, we have $x'\otimes
c\oplus x''\otimes c\lhd^w \lambda (y'\otimes c\oplus y''\otimes
c)$, or $x\otimes c\lhd^w\lambda y\otimes c$,  if and only if
${\rm min\ }y'\otimes c < {\rm max\ }y''\otimes c^m$, which is
equivalent to $g_u(c^m)<\frac{y_{d+1}}{y_d}$. The condition
$l_u(c^m)>g_u(y'')=\frac{y_n}{y_{d+1}}$ is automatically satisfied
by the assumption that $c^m\in [c]_{g_u(y'')}$. \hfill $\square$\\

The following lemma is a powerful tool in the study of the
mathematical structure of probabilistic MLOCC transformations.

\begin{lemma}\label{probboundary}\upshape
For any $x\in K_d^\lambda(y)$,  $x^{\otimes k}\lhd^w \lambda^k
y^{\otimes k}\Leftrightarrow y_{d+1}^k>y_n^{k-1}y_d.$
\end{lemma}

The proof of Lemma \ref{probboundary} is much more simpler than
its deterministic counterpart  Lemma \ref{uniqueinterior}. So we
prefer to give a proof here.

\textit{Proof} Let us decompose  $x$ and $y$ into $(x',x'')$ and
$(y',y'')$, respectively. By binomial theorem, we have
\begin{equation}\label{eq7}
(x^{\otimes k})^\downarrow=(b_x \oplus  x''^{\otimes
k})^\downarrow{\rm\ and\ } (y^{\otimes k})^\downarrow=(b_y \oplus
y''^{\otimes k})^\downarrow,
\end{equation}
where
$$b_x=\bigoplus_{i=0}^{k-1}(x'^{\otimes k-i}\otimes x''^{\otimes i})^{\oplus \choose{k}{i}} {\rm\ and\ }b_y=\bigoplus_{i=0}^{k-1}(y'^{\otimes k-i}\otimes  y''^{\otimes i})^{\oplus \choose{k}{i}}.$$
Since $x\in K_d^\lambda(y)$, it is obvious that $x'\lhd^w \lambda
y'$ and $x''\lhd \lambda y''$.  Applying Lemma \ref{strictsuper}
repeatedly, we have $b_x\lhd^w \lambda^k b_y$. By Corollary
\ref{corointerior}, we have  $x''^{\otimes k}\lhd \lambda^k
y''^{\otimes k}$. Hence by Lemma \ref{strictsuper} again, we
deduce from Eq.(\ref{eq7}) that  $x^{\otimes k}\lhd^w \lambda^k
y^{\otimes k}$ if and only if
$${\rm max\ }\lambda^ky''^{\otimes k}>{\rm min\ }\lambda^k b_y,$$
which is equivalent to $y_{d+1}^k>y_n^{k-1}y_d$. With that we
complete the
proof of the lemma.\hfill $\square$\\

Now we establish the following interesting result, which is
exactly a probabilistic version of Theorem \ref{minterior}.

\begin{theorem}\label{probminterior}\upshape
For any $y\in V^n$ and $0<d<n-1$, the following are equivalent:

1) $K_d^\lambda(y)\subseteq (M_k(y))^o$.

2) $K_d^\lambda(y)\subseteq (T_k(y))^o$.

3) $y_{d+1}^k>y_n^{k-1}y_d$.
\end{theorem}

\textit{Proof.} We first prove the equivalence of 2) and 3).
 Suppose 3) holds. We can choose
$c=(1,\alpha,\cdots,\alpha^{k-1})$ with $0<\alpha<1$ such that
$$(\frac{y_n}{y_{d+1}})^{k-1}<\alpha^{k-1}<\frac{y_{d+1}}{y_d}.$$
A routine calculation shows that $l_u(c)=\alpha{\rm \ and  \ }
g_u(c)=\alpha^{k-1}.$ Take $x\in K_d^\lambda(y)$. By the proof of
Theorem \ref{characterizationforcatalystusefulprob}, we have
$x\otimes c\lhd^w \lambda y\otimes c$. Moreover, we have
$K_d^\lambda (y)\subseteq (T^\lambda(y,c))^o\subseteq
(T_k^\lambda(y))^o.$

Conversely, $K_d^\lambda (y)\subseteq (T_k^\lambda(y))^o$ implies
that there exists $c\in V^k$ such that $S^\lambda (y)\subsetneqq
T^\lambda(y,c)$. By Theorem
\ref{characterizationforcatalystusefulprob}, we declare that there
exists $c'\in V^{k'}$ satisfying condition (\ref{probusecata}),
where $c'$ is a segment of $c$. By the relation $
l^{k'-1}_u(c')\leq g_u(c')$ and $k'\leq k$ we have
$l^{k-1}_u(c')\leq g_u(c')$. This fact together with condition
(\ref{probusecata}) shows that $y_{d+1}^k> y_dy_n^{k-1},$ which is
exactly the same as 3).

By Theorem \ref{mloccprobinterior}, the equivalence of 1) and 3)
is just to prove the fact: for any $x\in K_d^\lambda(y)$,
$x^{\otimes k}\lhd^w \lambda^k y^{\otimes k}$ if and only if
$y_{d+1}^k>y_n^{k-1}y_d$. This fact is guaranteed by
Lemma \ref{probboundary}. \hfill $\square$\\

Now we can present the  proof of Theorem \ref{kemloccprob} as
follows:

\vspace{1em} \textit{Proof of Theorem \ref{kemloccprob}.} Theorem
\ref{kemloccprob} is essentially implied by Theorem
\ref{probminterior}. To see this, we need only to prove that
$S^{\lambda}(y)\subsetneqq T_k^\lambda(y)$ and
$S^\lambda(y)\subsetneqq M_k^\lambda(y)$ are equivalent to
$K_d^\lambda(y)\subseteq (T_k^\lambda(y))^o$ and
$K_d^\lambda(y)\subseteq (M_k^\lambda(y))^o$ for some $0<d<n-1$,
respectively. The proofs are the same as that for the
deterministic case, since $S^\lambda(y)$, $T_k^\lambda(y)$ and
$M_k^\lambda(y)$ are all compact subsets of $V^n$. We omit the
details here, and thus complete the proof of  Theorem \ref{kemloccprob}.\hfill $\square$\\

 \section{conclusion}

To summarize, we have examined the powers of entanglement
catalysis and multiple-copy entanglement transformation  in three
different contexts: 1) the catalyst state is specified, 2) the
dimension of catalyst state is fixed, and  3) the number of copies
used in multiple-copy entanglement transformation is fixed. In the
case of 1), we have presented  an economic  sufficient condition
under which an entangled quantum state $c$ can serve as a catalyst
in producing the state $y$, i.e., $S(y)\subsetneqq T(y,c)$. In a
special case when $c$ has only two non-zero different Schmidt
coefficients, this condition is shown to be also a necessary one.
As an interesting application of this condition,  for any
nonuniform entangled state $z$ and $n\geq 4$, we can explicitly
construct a subset $A(z)$ of $V^n$, such that $z$ can catalyze any
states in $A(z)$. We have demonstrated that our result serves as
an extensively generalized version of the one obtained by  Daftuar
and Klimesh in \cite{DK01}, and thus is a more stronger answer to
Nielsen's conjecture \cite{MAJ}: any nonuniform entangled state
can serve as quantum catalyst for some entanglement
transformation.

In the cases of 2) and 3),  we have generalized the known result
$T(y)=S(y) \Leftrightarrow M(y)=S(y)$ to a finer one:
$T_k(y)=S(y)\Leftrightarrow M_k(y)=S(y)$. That is, $k$-ELOCC is
useful in producing $y$ if and only if $k$-MLOCC is useful in
producing $y$. Furthermore, an analytical condition for the
equality $T_k(y)=S(y)$ (and equivalently $M_k(y)=S(y)$) have been
found in terms of the components of $y$. We have also shown some
interesting applications of this result. Especially, for any
positive integer $k>1$, we have constructed a class of $4\times 4$
states which can be catalyzed by some catalysts with the dimension
at least $k+1$.  Our results can be generalized to probabilistic
transformations. Some differences between deterministic
transformations and probabilistic transformations have also been
discussed.

\smallskip\

\section*{Appendix A: Proof of Lemma \ref{interiorpointdirectsum}}
To be more specific, assume $y\in V^m$ and $y'\in V^n$.  Take
$x\lhd y$ and $x'\lhd y'$.

`$\Leftarrow$'. Suppose
\begin{equation}\label{overlap}
y_1>y_n' {\rm\ \ \ and\ \ } y_m<y_1'.
\end{equation} We  will prove  that $x\oplus x'$ is in the
interior of $S(y\oplus y')$. It suffices to show
\begin{equation}\label{dirmaj}
e_l(x\oplus x')<e_l(y\oplus y')
\end{equation}
for any $1\leq l< m+n$.

One can easily verify
\begin{equation}\label{chain2}
e_l(x\oplus x')=e_p(x)+e_q(x')\leq e_p(y)+e_q(y')\leq e_l(y\oplus
y'),
\end{equation}
where $0\leq p\leq m, 0\leq q\leq n$ and $p+q=l$. To complete the
proof, we  need to  consider the following two cases:

Case 1:  $0<p<m$ or $0<q<n$. By the conditions that $x\in S^o(y)$
and $x'\in S^{o}(y')$, we have
\begin{equation}
 e_p(x)<e_p(y)
{\rm \ or\ } e_q(x')<e_q(y').
\end{equation}
Then the first inequality in Eq.(\ref{chain2}) is strict, and
Eq.(\ref{dirmaj}) follows immediately.

Case 2: Either $p=m$ and $q=0$, or $p=0$ and $q=n$. They both
contradict the assumption in Eq.(\ref{overlap}). So we finish the
proof of the sufficiency part.

`$\Rightarrow$'. By contradiction, suppose that Eq.(\ref{dirmaj})
holds for very $1\leq l<m+n$ but Eq.(\ref{overlap}) does not hold.
If $y_1\leq y_n'$ then
\begin{equation}
e_n(y\oplus y')=e_n(y')=e_n(x')\leq e_n(x\oplus x'),
\end{equation}
a contradiction with Eq.(\ref{dirmaj}) when $l=n$. Similarly, if
$y_m\geq y_1'$ then
\begin{equation}
e_m(y\oplus y')=e_m(y)=e_m(x)\leq e_m(x\oplus
 x'),
\end{equation}
which contradicts  Eq.(\ref{dirmaj}) again. That completes the proof of the lemma.\hfill$\square$\\

\section*{Appendix B: Proof of Lemma \ref{tyinterior}}
We  need only to prove that under the constraint of
$l_u(c)>g_u(y)$, $x\in T^o(y,c)$ implies $x\otimes c\lhd y\otimes
c$. By contradiction, suppose that for some $x\in T^o(y,c)$, it
holds that $x\otimes c\notin S^o(y\otimes c)$. Then there should
exist some $1\leq l\leq nk-1$ such that $e_l(x\otimes
c)=e_l(y\otimes c).$ However, as we will show latter, this case is
impossible.

First,  we  prove that $e_l(x\otimes c)=e_l(y\otimes c)$ and
$1\leq l\leq nk-1$ imply  $l=i_0n$ for some $1\leq i_0<k$. For
this purpose, let us define $$\bar{x}=(x_1+\epsilon,\cdots,
x_n-\epsilon).$$ It is obvious that $x\lhd \bar{x}$ for any
$\epsilon>0$. Since $x\in T^o(y,c)$,  we have $\bar{x}\in
T^o(y,c)$ for a sufficiently small positive real $\epsilon$, which
implies $\bar{x}\otimes c\prec y\otimes c$. Then  we have
$e_l(\bar{x}\otimes c)\leq e_l(y\otimes c)$. Noticing
$e_l(x\otimes c)\leq e_l(\bar{x}\otimes c)$ and the assumption
$e_l(x\otimes c)=e_l(y\otimes c)$, we have $e_l(x\otimes
c)=e_l(\bar{x}\otimes c)$.  Choose $l_1,\cdots, l_k$ such that
$$e_l(x\otimes c)=\sum_{i=1}^k c_ie_{l_i}(x),$$
where $\sum_{i=1}^k l_i=l$ and $0\leq l_i\leq n$. If there exists
$1\leq p\leq k$ such that $0<l_{p}<n$, then by $x\lhd \bar{x}$ we
have $e_{l_p}(x)<e_{l_p}(\bar{x})$. Thus $$e_l(x\otimes
c)=\sum_{i=1}^k c_ie_{l_i}(x)<\sum_{i=1}^k c_ie_{l_i}(\bar{x})\leq
e_l(\bar{x}\otimes c),$$ which contradicts $e_l(x\otimes
c)=e_l(\bar{x}\otimes c)$. Therefore, for any $1\leq i\leq k$, we
have $l_i\in \{0,n\}$. Taking $i_0={\rm max}\{i:l_i>0\}$, we have
$1\leq i_0<k$ by the assumption $1\leq l\leq nk-1$. So, $l=i_0n$.

Second, we show $e_l(x\otimes c)<e_l(y\otimes c)$ for $l=i_0n$. In
fact, by the above argument, we have
\begin{equation}\label{eq1}
e_{i_0n}(x\otimes
c)=\sum_{i=1}^{i_0}c_ie_n(x)=\sum_{i=1}^{i_0}c_ie_n(y).
\end{equation} Notice that $l_u(c)>g_u(y)$ yields
$c_{i_0+1}y_1>c_{i_0}y_n$. So
\begin{equation}\label{eq2}
\sum_{i=1}^{i_0}c_ie_n(y)=\sum_{i=1}^{i_0-1}c_ie_n(y)+c_{i_0}e_{n-1}(y)+c_{i_0}y_n<\sum_{i=1}^{i_0-1}c_ie_n(y)+c_{i_0}e_{n-1}(y)+c_{i_0+1}y_1\leq
e_{i_0n}(y\otimes c),
\end{equation}
where the last inequality is by the definition of
$e_{i_0n}(y\otimes c)$. Combining Eqs.(\ref{eq1}) with (\ref{eq2})
shows $e_l(x\otimes c)<e_l(y\otimes c)$ for $l={i_0n}$, again a
contradiction. With that we complete the proof of the lemma.
\hfill $\square$\\

\section*{Appendix C:  Proof of Lemma \ref{uniqueinterior}}
Let $x'=(x_1,\cdots,x_d)$ be the vector formed
 by the $d$ largest components of $x$, and $x''$ is the rest part of
 $x$. We can similarly define $y'$ and $y''$. Then it is easy to check
 \begin{equation}\label{2interior}
        x'\lhd y' {\rm \ and\ } x''\lhd y''
\end{equation}
by the definition of $K_d(y)$.  Also we have
\begin{equation}\label{decom2}
x=x' \oplus x''{\rm \ and\ } y=y' \oplus y''.
\end{equation}

 We give a  proof of the part `$\Leftarrow$' by seeking a
 sufficient condition for $x^{\otimes k}\lhd y^{\otimes k}$.
 First we notice the following identity by binomial
 theorem:
\begin{equation}\label{tensorbionormial}
  (y^{\otimes k})^{\downarrow}=(\bigoplus_{i=0}^{k} (y'^{\otimes (k-i)}\otimes y''^{\otimes
  i})^{\oplus \choose{k}{i}})^{\downarrow}.
\end{equation}
And $x^{\otimes k}$ has a similar expression.  For simplicity, we
denote
\begin{equation}\label{yicon}
y^i=(y'^{\otimes (k-i)}\otimes y''^{\otimes
  i})^{\downarrow}, n_i=d^{k-i}(n-d)^i.
\end{equation}
And $x^i$ is defined similarly.

     Noticing Eqs.(\ref{2interior}) and (\ref{yicon}), we have
\begin{equation}
    x^i\lhd y^i {\rm\ for\ any \ }0\leq i\leq k
\end{equation}
 by
Corollary \ref{corointerior}. So to ensure $x^{\otimes k}\lhd
y^{\otimes k}$, we only need that the set $A=\{(y^i)^{\oplus
\choose{k}{i}}:0\leq i\leq k\}$ satisfies the conditions in 2) of
Corollary \ref{sumrepeatset}. It is easy to check that
$${\rm max\ }y^0={\rm max}\{{\rm max\ }y^i:0\leq i\leq k \}{\rm\
and\ } {\rm min\ }y^k={\rm min}\{{\rm min\ }y^i:0\leq i \leq
k\}.$$ Hence we only need $A$ to satisfy the overlapping
conditions, i.e., ${\rm min\ }y^i<{\rm max\ }y^{i+1}$, or more
explicitly,
\begin{equation}\label{overlapcondition}
                y_d^{k-i}y_n^i<y_1^{k-(i+1)}y_{d+1}^{i+1}, {\rm \ \ }0\leq
                i<k.
\end{equation}
By the monotonicity,  Eq.(\ref{overlapcondition}) is just
equivalent to the cases of $i=0$ and $i=k-1$. That is,
\begin{equation}\label{simpleoverlapcondition}
                y_d^k<y_1^{k-1}y_{d+1} {\rm \ and\ }
                y_{d+1}^k>y _n^{k-1}y_d,
\end{equation}
which is exactly the condition in  the right-hand side of
Eq.(\ref{2interiorcondition}). That completes  the proof of the
part `$\Leftarrow$'.

     Now we prove the part `$\Rightarrow$'. By contradiction,
suppose the conditions in the right-hand side of
Eq.(\ref{2interiorcondition}) are satisfied. Then there should
exist $0\leq i_0<k$ that violates the conditions in
Eq.(\ref{overlapcondition}), i.e.,
\begin{equation}\label{violateoverlapcondition}
                y_d^{k-i_0}y_n^{i_0}\geq
                y_1^{k-(i_0+1)}y_{d+1}^{i_0+1}.
\end{equation}
But then we can deduce that
\begin{equation}
  e_{d(i_0)}(y^{\otimes
  k})=\sum_{i=0}^{i_0}\choose{k}{i}e_{n_i}(y^i)=\sum_{i=0}^{i_0}\choose{k}{i}e_{n_i}(x^i)\leq e_{d(i_0)}(x^{\otimes
  k}),
\end{equation}
which contradicts  the assumption $x^{\otimes k}\lhd y^{\otimes
k}$, where $d(i_0)=\sum_{i=0}^{i_0}\choose{k}{i}n_i$.

   With that we complete the proof of the lemma.\hfill$\square$\\

\end{document}